\documentclass{ws-ijmpa}
\usepackage{graphicx,graphics,epsfig,pifont}
\usepackage{latexsym}
\usepackage{txfonts}
\usepackage{hyperref}
\usepackage[english]{babel}
\RequirePackage[T1]{fontenc}
\RequirePackage[numbers]{natbib}
\usepackage{mathptmx}      
\usepackage{amssymb}      
\usepackage{amsmath}

\begin{document}

\title{ General recipe to form input space for deep learning analysis of HEP scattering processes. }

\author{ Lev Dudko \footnote{lev.dudko@cern.ch}, Georgi Vorotnikov , Petr Volkov , Maxim Perfilov, }
\address{Skobeltsyn Institute of Nuclear Physics of Lomonosov Moscow State University (SINP MSU), 1(2), Leninskie gory, GSP-1, Moscow 119991, Russian Federation }
\author{ Andrei Chernoded, Dmitri Ovchinnikov, Artem Shporin}
\address{Faculty of Physics, Lomonosov Moscow State University, Leninskie Gory, Moscow 119991, Russian Federation}

\maketitle

\begin{abstract}
Deep learning neural network technique (DNN) is one of the most efficient and general approach of multivariate data analysis of the collider experiments. The important step of the analysis is the optimization of the input space for multivariate technique. In the article we propose the general recipe how to form the set of low-level observables sensitive to the differences in hard scattering processes at the colliders.   
It is shown in the paper that without any sophisticated analysis of the kinematic properties one can achieve close to optimal performance of DNN with the proposed general set of low-level observables. 
\end{abstract}

%


\section*{Introduction}
\label{intro}
Many high energy physics tasks in the collider experiments require modern efficient techniques to reach the desired sensitivity. Rather general scheme of HEP data analysis contains the distinguishing of some rare signal physics process from overwhelming background processes. Neural network technique (NN) is one of the most popular and efficient multivariate method to analyze multidimensional space of the observables. NN is powerful approach to increase sensitivity of the experiment. Possible optimizations of the set of high-level observables for multivariate analysis were considered previously, and general recipe was formulated~\cite{LevDudko:1999gna,Boos:2003gv,Boos:2008sdz} based on the analysis of Feynman diagrams which contribute to signal and background processes. The novel approach of deep learning neural network (DNN) becomes more popular and most efficient in some cases~\cite{Baldi:2014kfa}. The main advantage of DNN is the ability to operate with raw low-level unpreprocessed data and recognition of the necessary features during the training stage. Unfortunately, the checks of naive implementation of DNN with low-level observables, such as four momenta of the final particles, do not demonstrate desired efficiency. The matter of this article is the understanding possible general set of low-level observables to search for the hard scattering processes in the collider experiments and provide the general recipe how to form the input space for such DNN analysis.

\section{Check of the conception}
\label{nn-functions}
\begin{figure}[!h!]
  \centering
\begin{minipage}[!h!]{.49\linewidth}
\includegraphics[width=.9\linewidth]{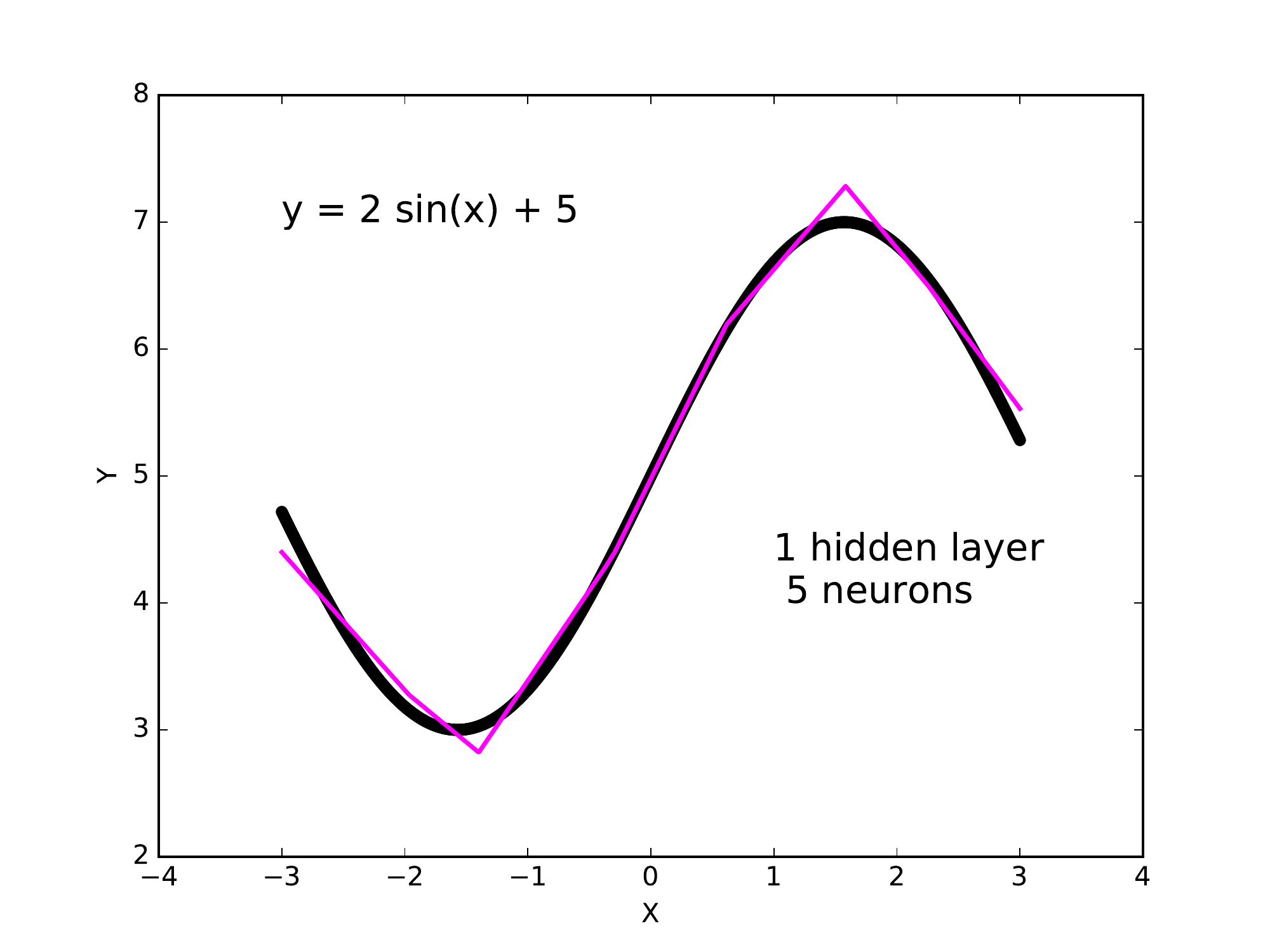}
\end{minipage}
\begin{minipage}[!h!]{.49\linewidth}
\includegraphics[width=.9\linewidth]{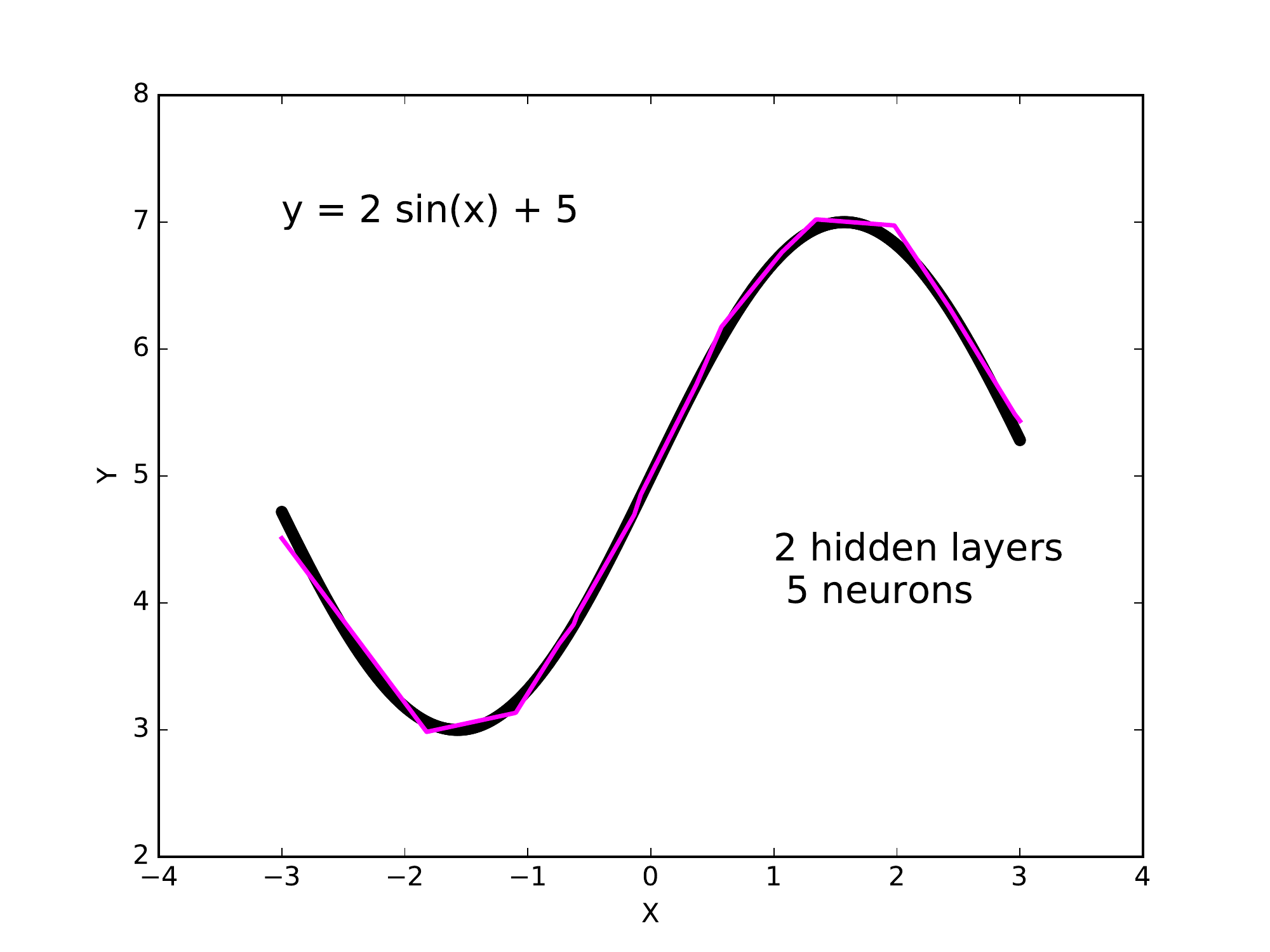}
\end{minipage}
\begin{minipage}[!h!]{.49\linewidth}
\includegraphics[width=.9\linewidth]{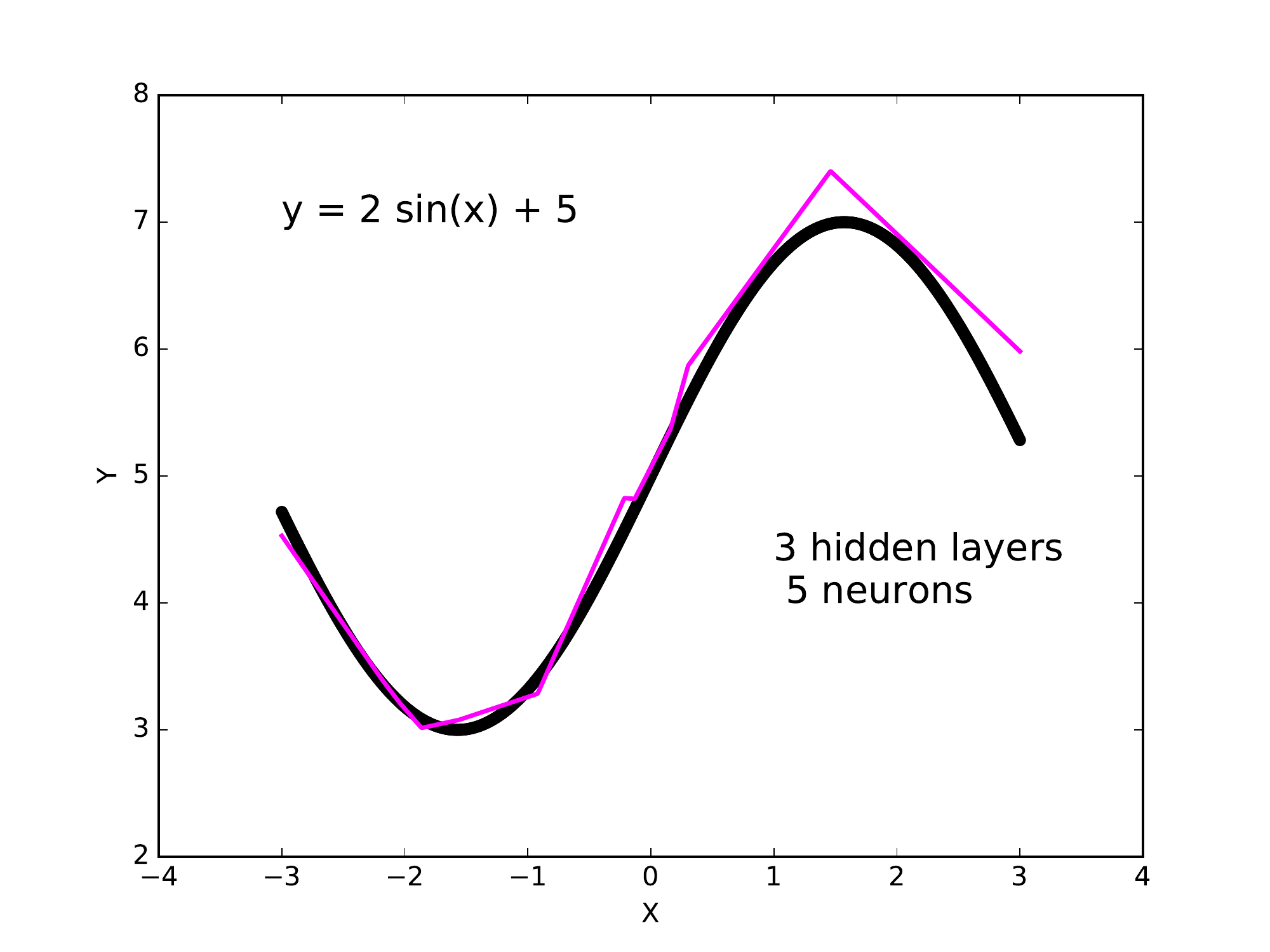}
\end{minipage}
\begin{minipage}[!h!]{.49\linewidth}
\includegraphics[width=.9\linewidth]{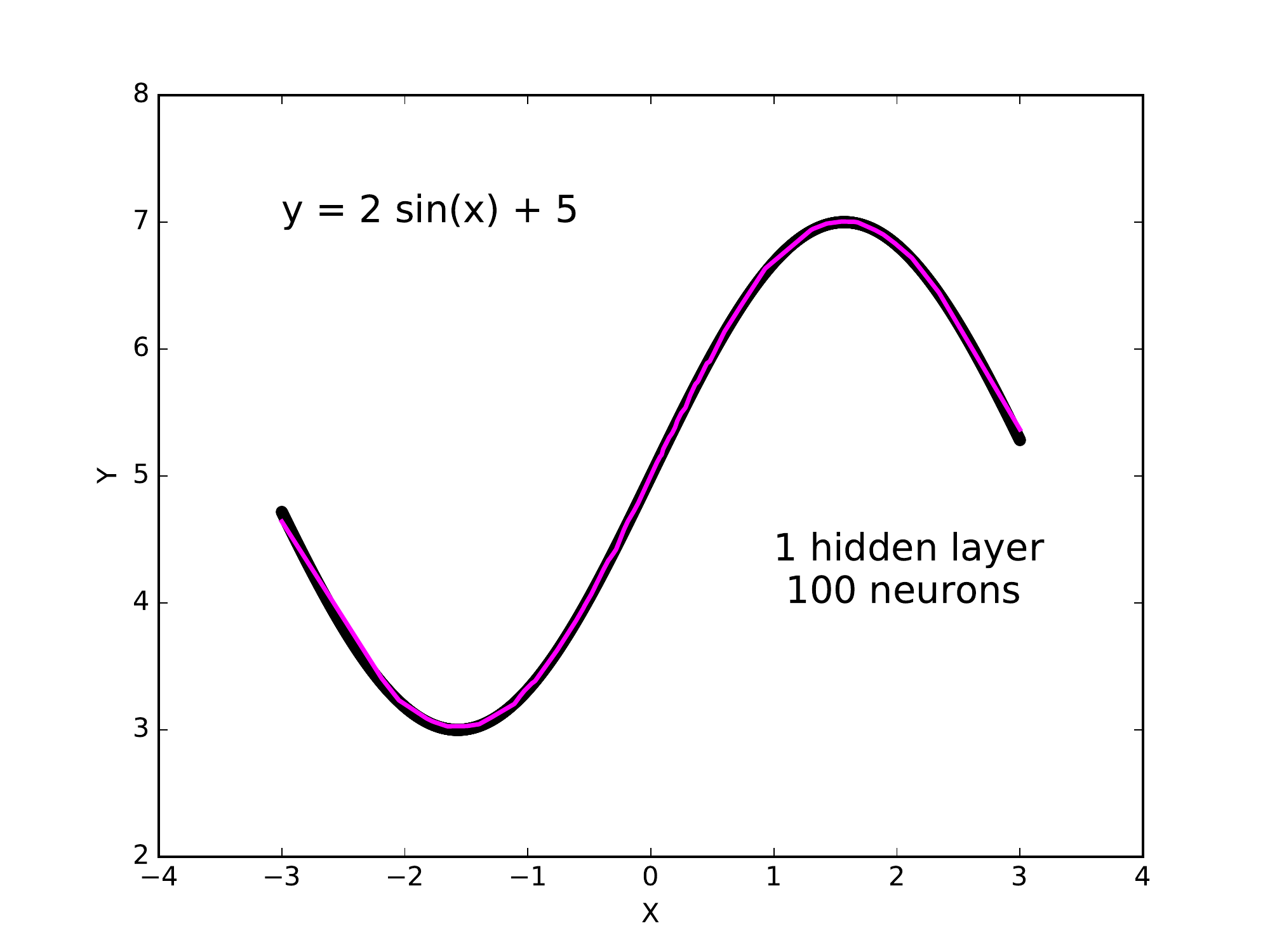}
\end{minipage}
\centering
\caption{\label{simple-function} Simple example of the approximation of non-linear function by DNN with ReLU activation function. }
\end{figure}

Training of the NN usually means an approximation of some function. The classification tasks can be considered as an approximation of the multidimensional function to match the class of input vector to the desired output of NN for this class. What types of functions can be approximated in this manner? The question can be traced historically to the 13th mathematical problem formulated by David Hilbert~\cite{Hilbert1,Hilbert2}. It can be formulated in the following way: ``can every continuous function of three variables be expressed as a composition of finitely many continuous functions of two variables?''. The general answer has been given by Andrey Kolmogorov and Vladimir Arnold~\cite{Kolmogorov,Arnold} in the Kolmogorov–Arnold representation theorem: ``every multivariate continuous function can be represented as a superposition of continuous functions of one variable''. Based on this theorem, one can conclude that the methods developed for NN training, potentially can approximate all continuous multivariate functions. In reality, if we consider the standard form of perceptron $y_i = \sigma (\sum_{j=1}^{n}w_{ij}x_j + \theta_i)$ the only nonlinear part is the activation function $\sigma()$. In the simplest case, if we take very popular activation function ReLU ($\sigma(x)=x$ for $x>0$ and $\sigma(x)=0$ for $x\le 0$), the whole NN with many layers and perceptrons is the linear combination of the input vector ($\bf x$). Very simple example of the approximation of nonlinear function $y=2\sin(x) + 5$ with DNN is shown in Fig.~\ref{simple-function}. ReLU activation function is taken for the simplicity.  The first plot in Fig.~\ref{simple-function} demonstrates 5 linear segments for the DNN output with 5 neuron on one hidden layer. Two hidden layers demonstrate the same linear segmentation of the DNN output. Three hidden layers make the approximation worse.  But the DNN with significantly increased number of neurons at one hidden layer provides the best smooth approximation of the non-linear function. 

The simplest general task for NN is an approximation of some non-linear function by linear segments, possible non-linearity in NN comes from nonlinear activation function. The representation theorem proves the possibility to approximate all continues functions, but in case the function is more complex than just linear function one needs to increase the complexity of the NN, increase number of hidden nodes and/or layers. In turn, the complexity complicates the training and the first possible optimization to achieve better performance is the simplifying of the training by decreasing the order of non-linearity in the objective function (desired DNN output depending on the DNN input variables).
Therefore, one need to understand what type of functions consider in High energy physics to describe the properties of the collider hard scattering processes and how one can describe the input space in most general and efficient way with the lowest order of non-linearity. 

In the collider physics the hard process properties are formulated based on four-momenta of the final particles. The reasonable way is to take all possible four-momenta of all final particles as the input vector for the DNN analysis and train DNN to distinguish sensitive features. We can try to check the hypothesis and find the general low-level  observables for the DNN implementation. For the DNN training we use Tensorflow~\cite{Abadi:2016kic} and Keras~\cite{chollet2015keras} software. Feed-Forward NN is used as an architecture of the DNN with forty to sixty hidden nodes on the three to five hidden layers depending on the number of inputs. Adam optimizer and dropout algorithm with 20\% of the exclusion level are used for the training. The standard overall (not batch) normalization of the inputs has been performed. The ReLU activation function was used for the intermediate layers and Sigmoid for the final layer. Variation of the architecture and training parameters demonstrate stable behavior for the considered tasks and particular configurations are specified below.

\begin{figure}[!h!]
\begin{minipage}[!h!]{.49\linewidth}
  \centering
\includegraphics[width=.95\linewidth]{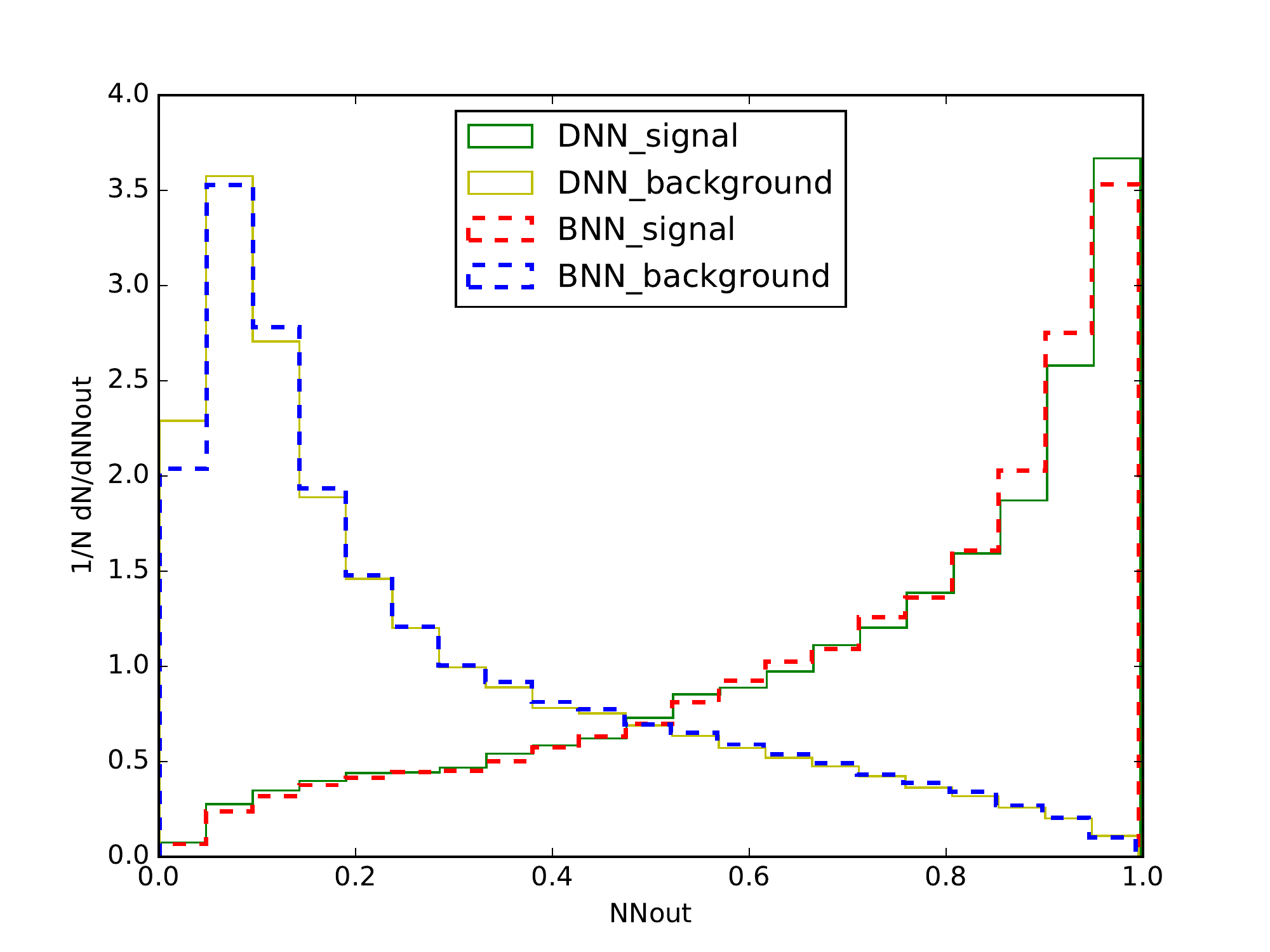}
\end{minipage}
\begin{minipage}[!h!]{.49\linewidth}
  \centering
\includegraphics[width=.95\linewidth]{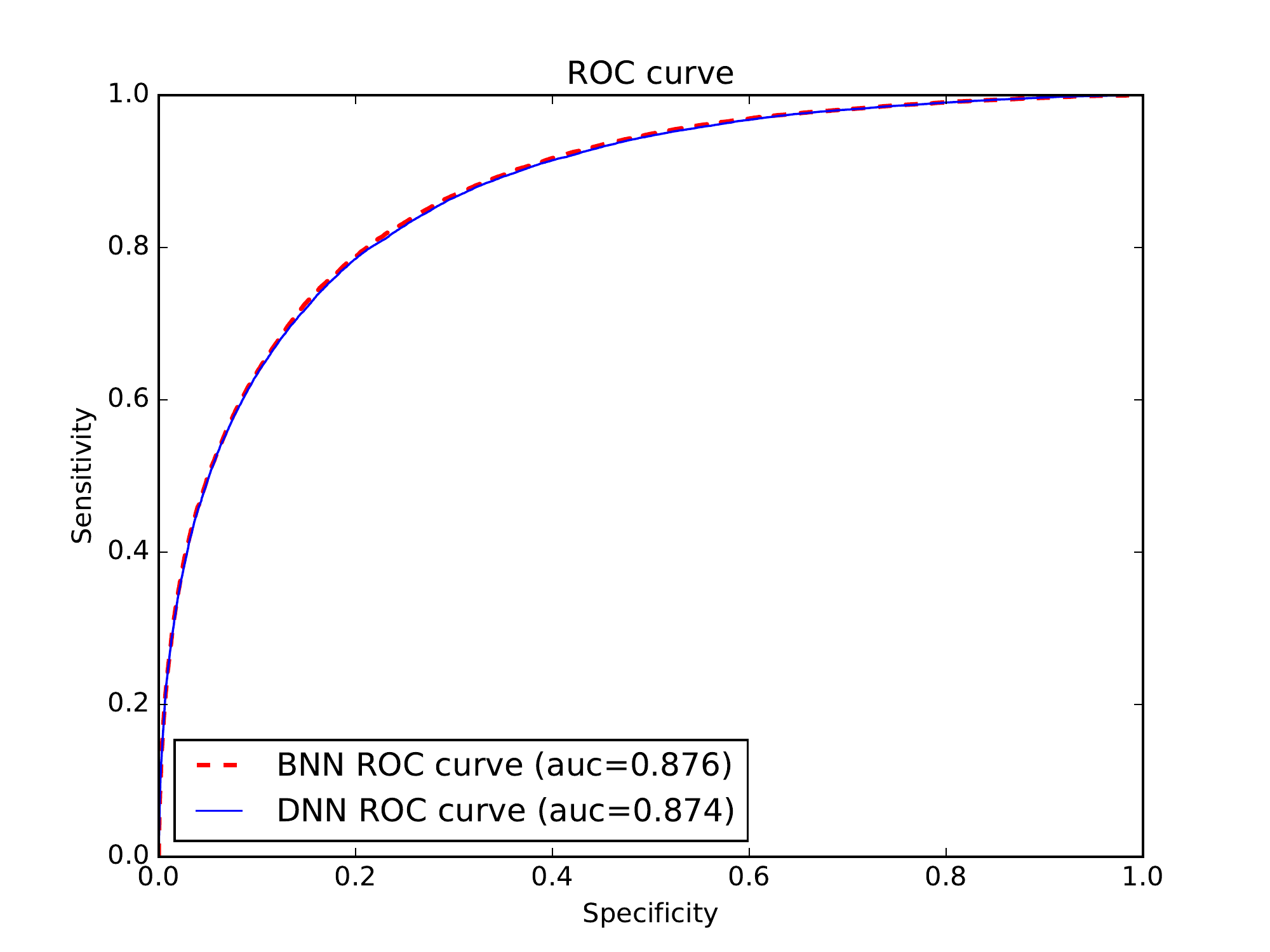}
\end{minipage}
\centering
\caption{\label{high-level-1layer} Comparison of the BNN and DNN trained on the same set of high-level input variables. BNN and DNN have one hidden layer. The left plot demonstrates outputs of DNN and BNN for the signal and background processes. The ROC curves are shown in the right plot.}
\end{figure}

For the criteria of the efficiency one can use more simple, but efficient Bayesian neural networks (BNN)~\cite{FBMBook,FBMPackage} with only one hidden layer. The set of high-level input variables for BNN is very specific for the particular physics processes. For the check we can use highly optimized set of high-level observables found by the method mentioned above~\cite{Boos:2008sdz}. As an example of the physics task we consider distinguishing of the t-channel single-top-quark production from the pair-top-quark production processes. The task is not trivial, but it is already considered many times in the past~\cite{Abbott:1999te}. The set of high-level input variables for the cross-check of the efficiency is the same as in the analysis of CMS collaboration~\cite{Khachatryan:2016sib}. The Monte-Carlo simulation of the signal and background events has been performed in CompHEP package~\cite{Boos:2004kh} with subsequent simulations of the detector response in PYTHIA~\cite{Sjostrand:2014zea} and DELPHES~\cite{deFavereau:2013fsa}. Simulation of the detector response does not add any additional kinematic properties of the processes, but it smears the existing differences between signal and background. The benchmark BNN and different DNNs are trained with the same simulated events, therefore the direct comparison of the BNN and DNN efficiencies  and achieved conclusions do not depend on the quality of the simulation of the detector. The training has been performed with about 180 thousands of simulated signal and background events with about 11 thousands of free parameters of the DNN. Separate event samples were used to estimate the efficiency of DNN to avoid overfitting problem, in additional to dropout algorithm.  
At the first step of the check we compare the efficiency of BNN and DNN with the same set of high-level input variables. The comparison is shown in the Fig.~\ref{high-level-1layer}. The left plot demonstrates output of DNN and BNN for the signal and background processes. The right plot demonstrates ROC (Receiver Operating Characteristics) curve which is usually used to demonstrate the efficiency. The efficiency is higher if the Area Under the Curve (AUC) is higher. 

In the Fig.~\ref{high-level-1layer} one can see the same efficiency of the BNN and DNN with one hidden layer, for the same set of high-level variables. We can conclude that both of the methods provide the same sensitivity for the same input vector. But, the DNN technique is able to analyze raw low-level information and prepare very large networks with many hidden layers to distinguish the sensitive features from low-level observables. 

\begin{figure}[!h!]
\begin{minipage}[!h!]{.49\linewidth}
  \centering
\includegraphics[width=.95\linewidth]{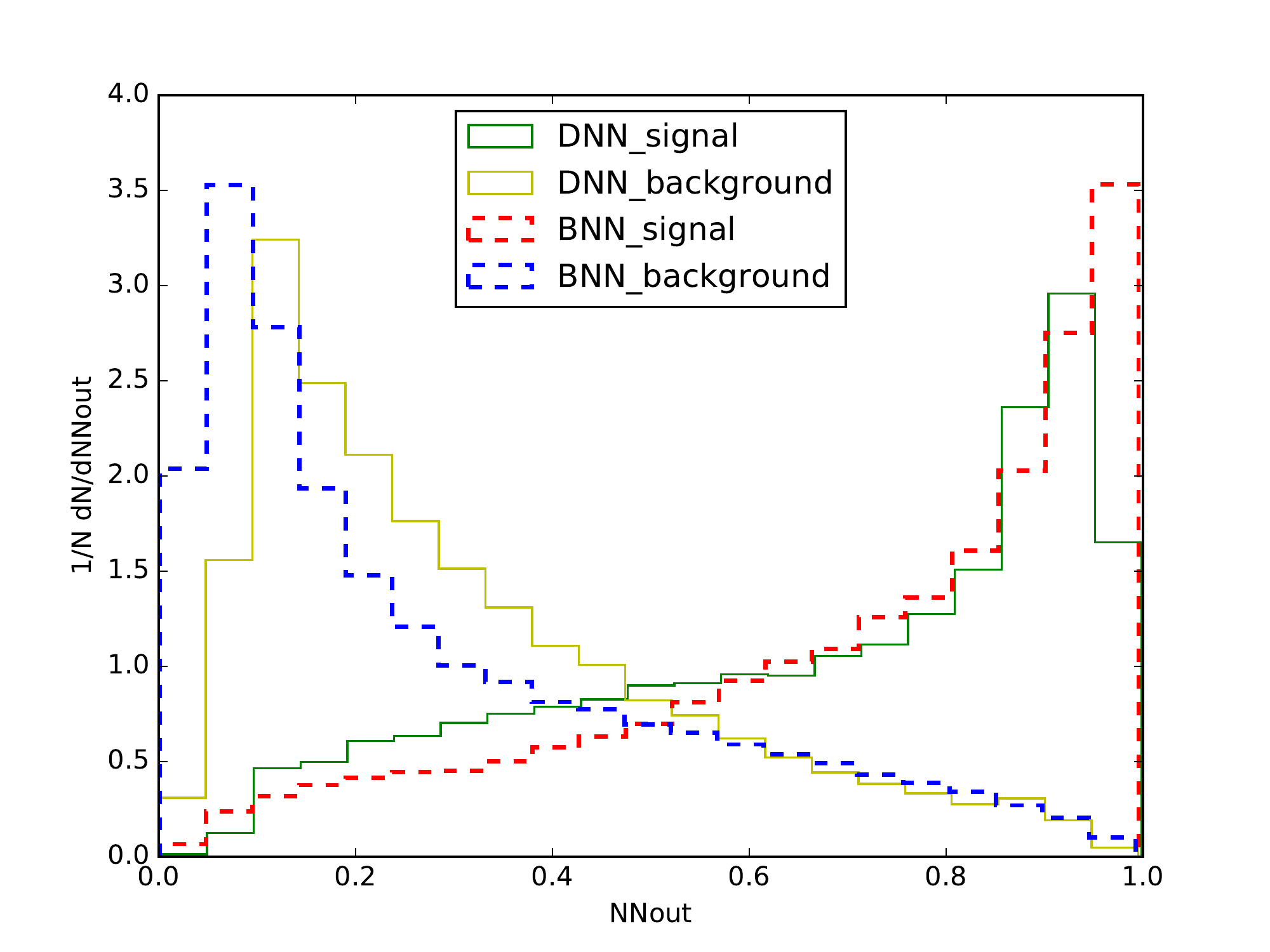}
\end{minipage}
\begin{minipage}[!h!]{.49\linewidth}
  \centering
\includegraphics[width=.95\linewidth]{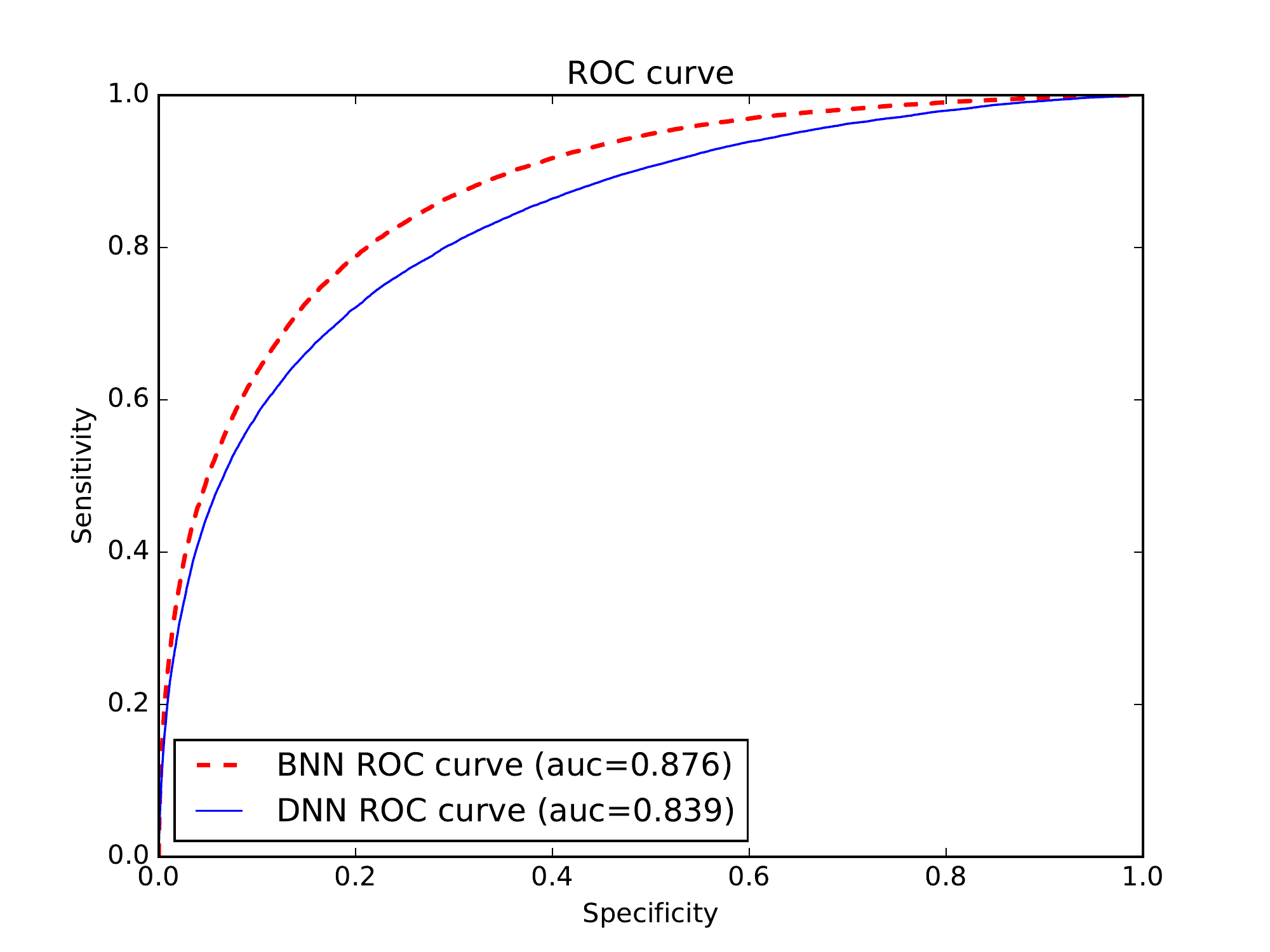}
\end{minipage}
\centering
\caption{\label{four-mom} The comparison of benchmark BNN with DNN trained on the naive set of input variables with four-momenta of the final particles. DNN has three hidden layers. The left plot demonstrates outputs of DNN and BNN for the signal and background processes. The ROC curves are shown in the right plot.}
\end{figure}

At the second step we check the naive set of low-level observables and compare the same benchmark BNN from the first step with DNN trained on the four-momenta of the final particles as an input vector. The corresponding comparison is shown in Fig.~\ref{four-mom}. One can see (Fig.~\ref{four-mom}) that DNN trained with four-momenta of the final particles provides significantly worse result than benchmark BNN trained with optimized high-level variables. Such behavior demonstrates that one need to understand deeper the function which has to be approximated and optimize the raw input vector to describe the input space for DNN.  Probably, it is also possible to achieve desired performance by significantly increased complexity of DNN (number of hidden nodes and/or layers), similarly as it is demonstrated in Fig~\ref{simple-function}. However, the increased complexity of DNN leads to more difficult training and is limited by the size of training sample. Therefore, it is not always possible to skip optimization steps and just increase the complexity of DNN.

\section{Formulation of the general recipe to form the input space for DNN analysis of HEP scattering processes.}
\label{recipe}

\begin{figure}[!h!]
\begin{minipage}[!h!]{.49\linewidth}
  \centering
\includegraphics[width=.95\linewidth]{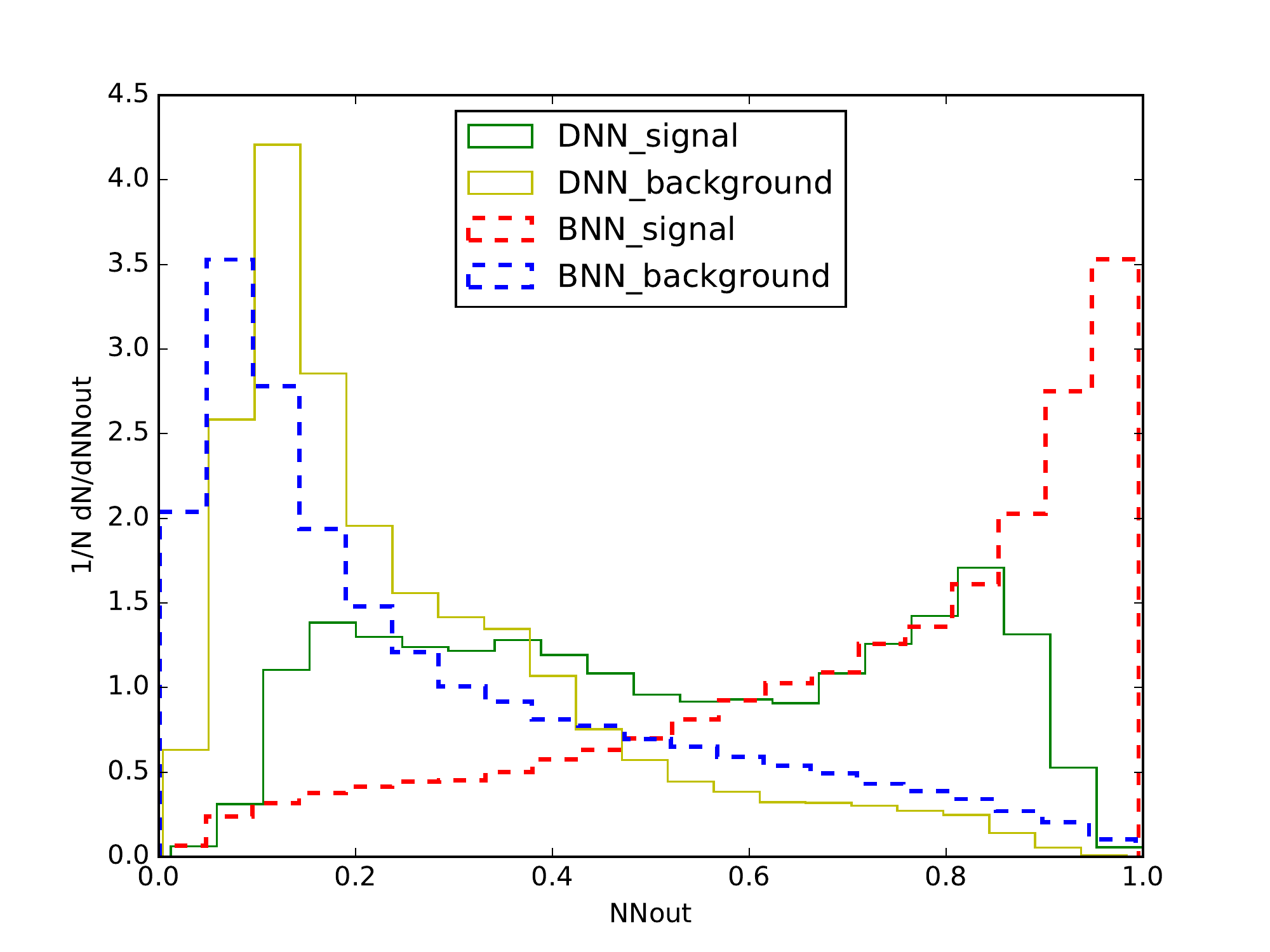}
\end{minipage}
\begin{minipage}[!h!]{.49\linewidth}
  \centering
\includegraphics[width=.95\linewidth]{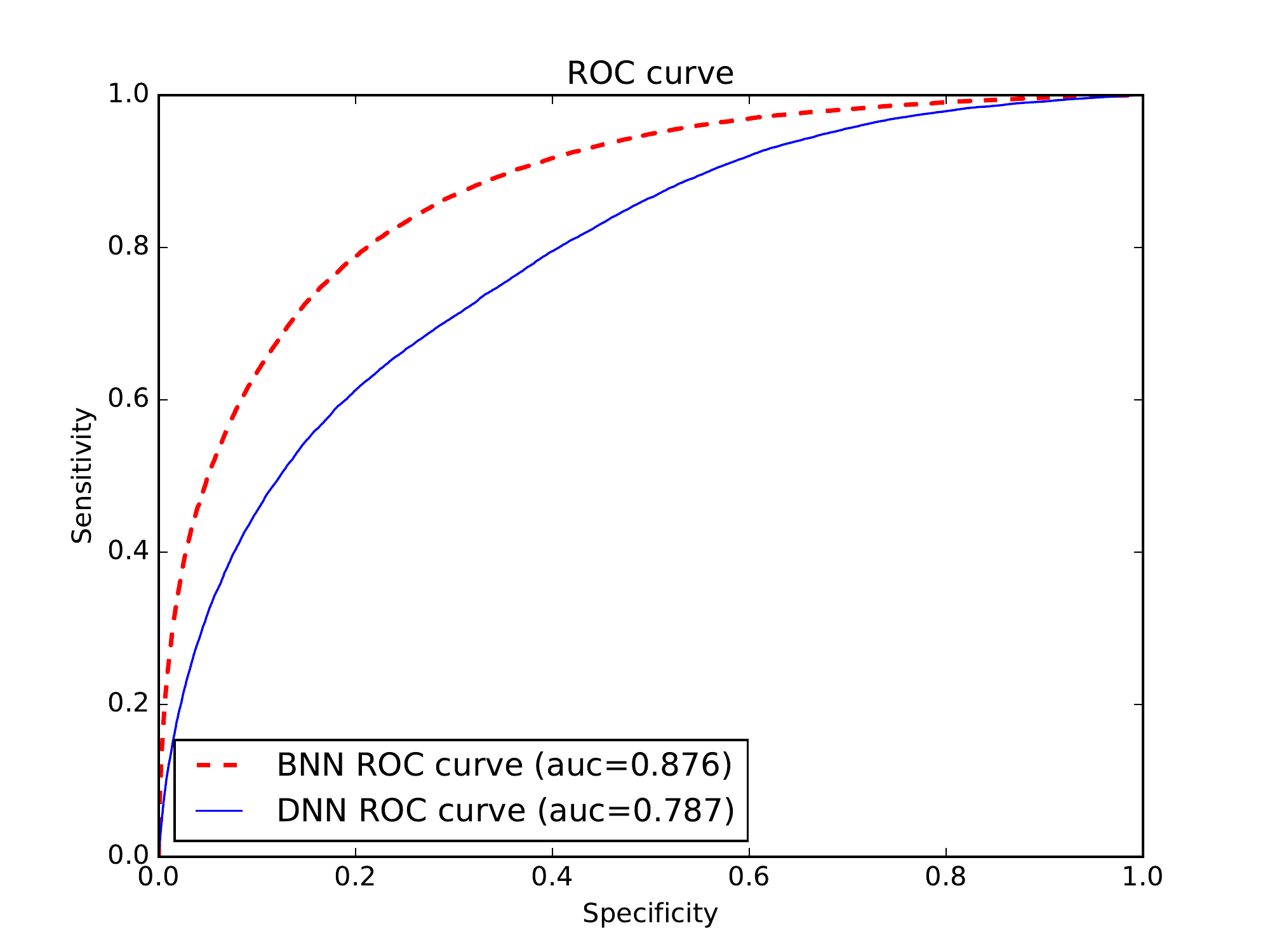}
\end{minipage}
\centering
\caption{\label{sp-3layers} The comparison of benchmark BNN with DNN trained on the scalar-products of four-momenta of the final particles as the set of input variables. DNN has three hidden layers. The left plot demonstrates outputs of DNN and BNN for the signal and background processes. The ROC curves are shown in the right plot.}
\end{figure}

\begin{figure}[!h!]
\begin{minipage}[!h!]{.49\linewidth}
  \centering
\includegraphics[width=.95\linewidth]{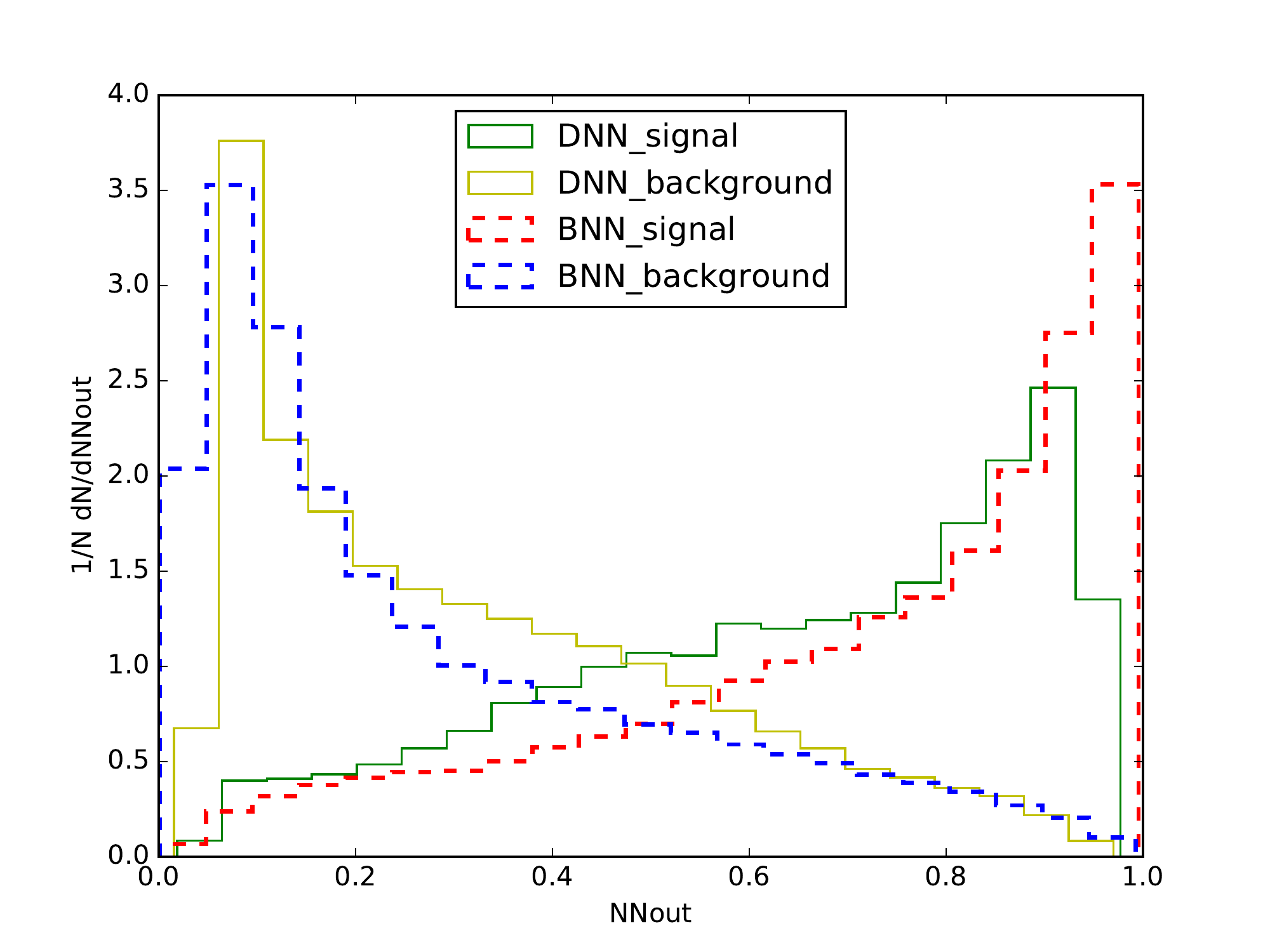}
\end{minipage}
\begin{minipage}[!h!]{.49\linewidth}
  \centering
\includegraphics[width=.95\linewidth]{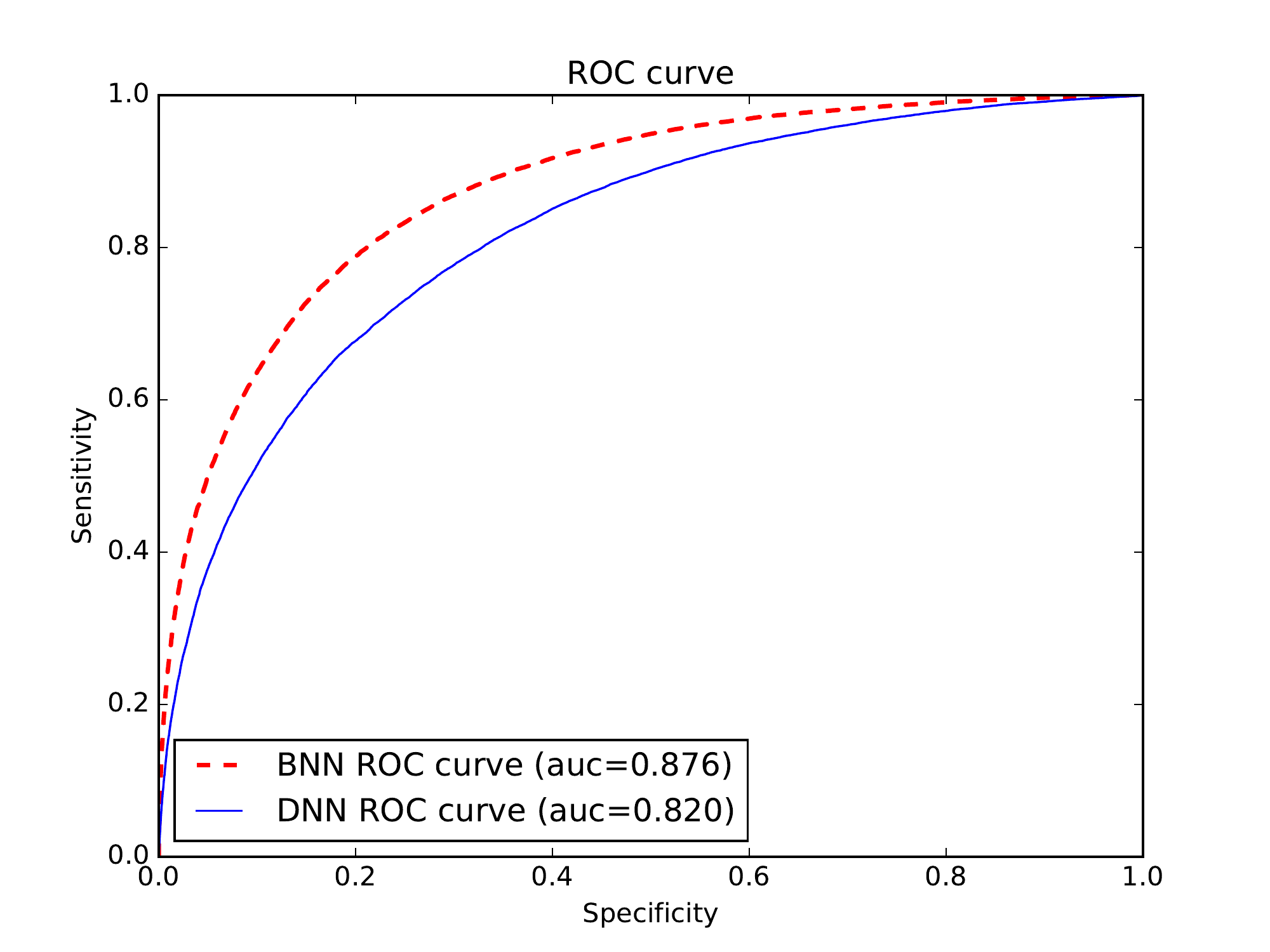}
\end{minipage}
\centering
\caption{\label{sp_mand_5layer} The comparison of benchmark BNN with DNN trained on the scalar-products of four-momenta of the final particles and s-channel Mandelstam variables as the set of input variables. DNN has five hidden layers. The left plot demonstrates outputs of DNN and BNN for the signal and background processes. The ROC curves are shown in the right plot.}
\end{figure}

\begin{figure}[!h!]
\begin{minipage}[!h!]{.49\linewidth}
  \centering
\includegraphics[width=.95\linewidth]{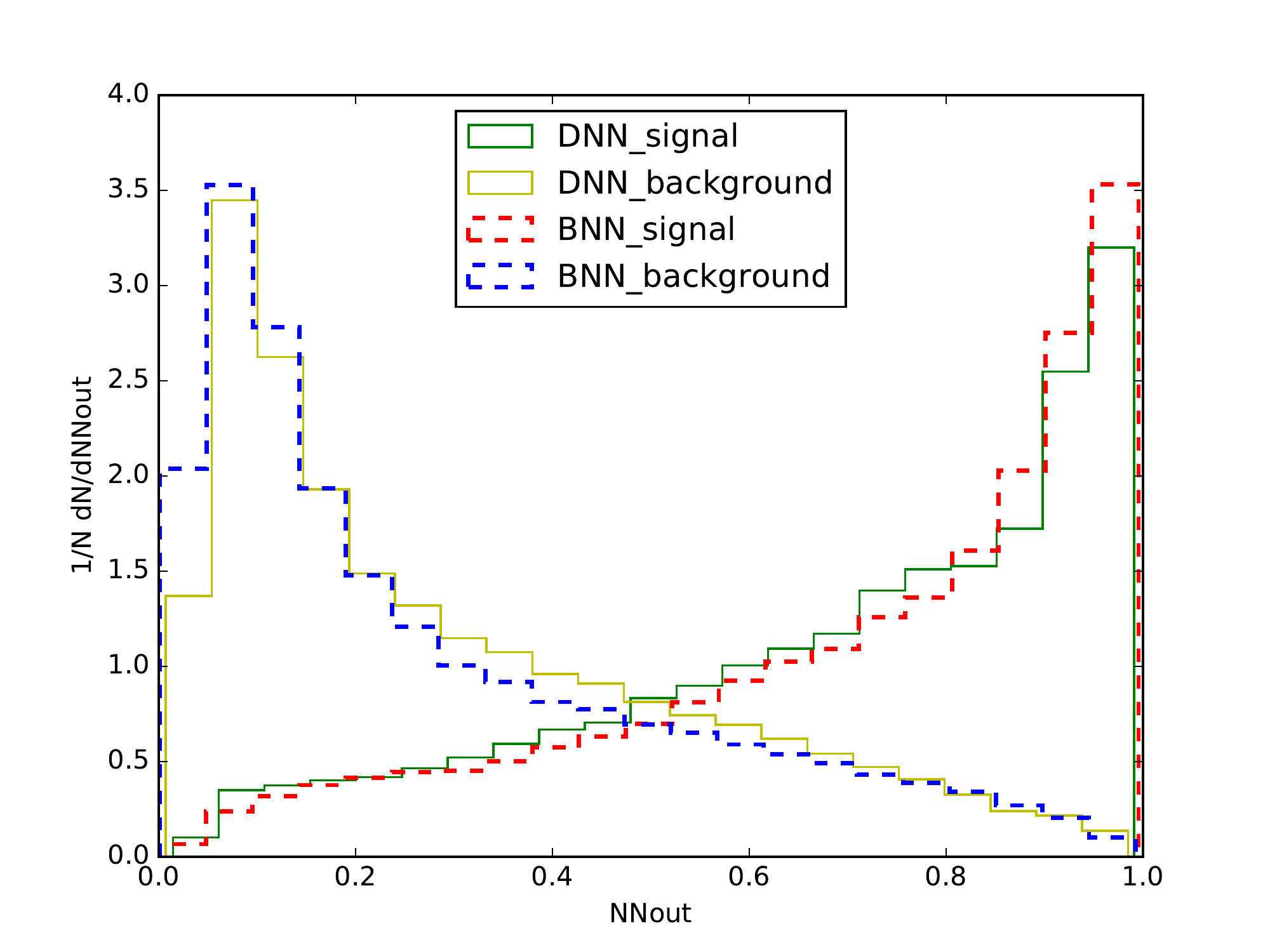}
\end{minipage}
\begin{minipage}[!h!]{.49\linewidth}
  \centering
\includegraphics[width=.95\linewidth]{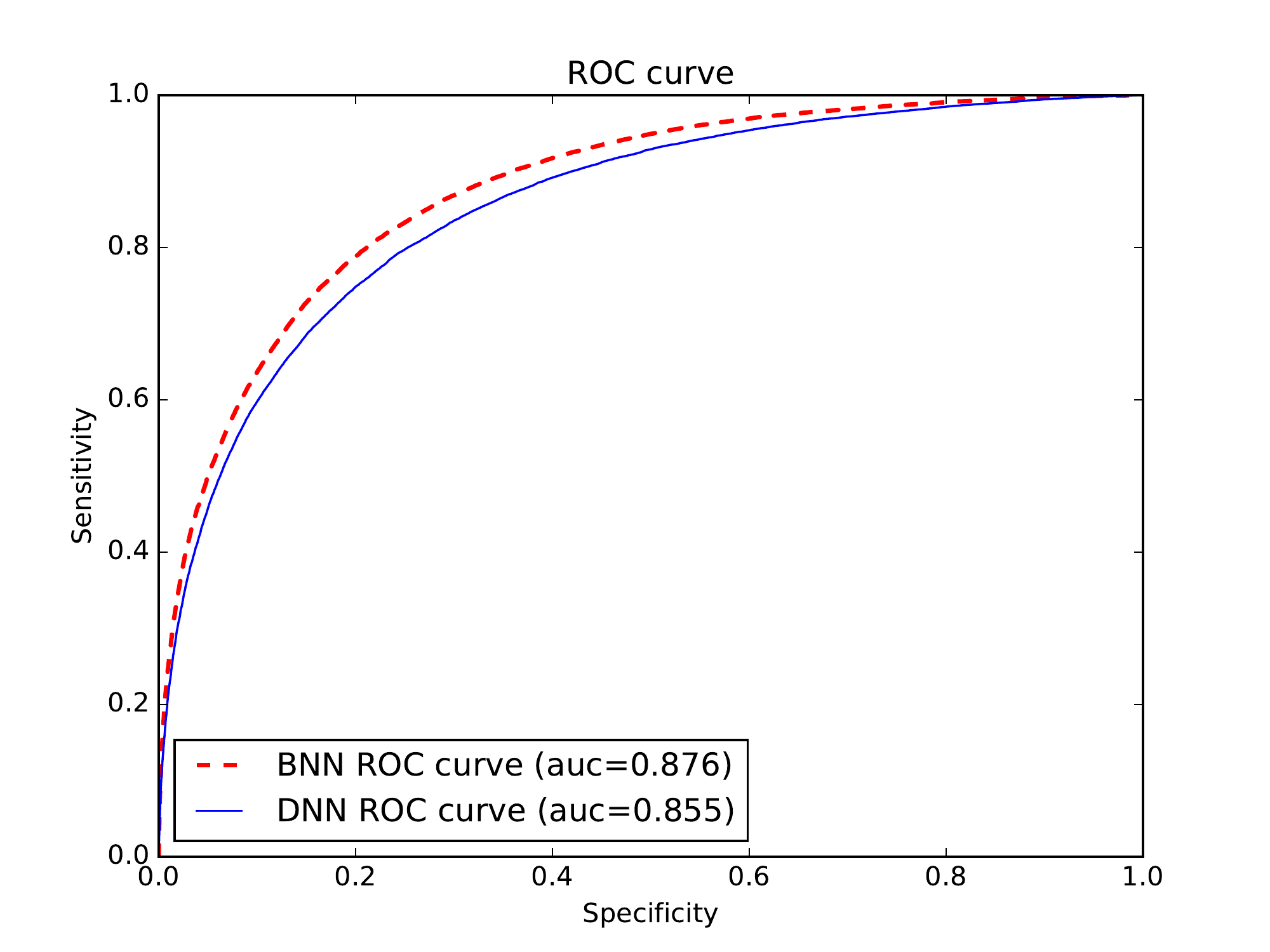}
\end{minipage}
\centering
\caption{\label{low_sp_5layer} The comparison of benchmark BNN with DNN trained on the scalar-products of four-momenta and four-momenta of the final particles as the set of input variables. DNN has five hidden layers. The left plot demonstrates outputs of DNN and BNN for the signal and background processes. The ROC curves are shown in the right plot.}
\end{figure}

\begin{figure}[!h!]
\begin{minipage}[!h!]{.49\linewidth}
  \centering
\includegraphics[width=.95\linewidth]{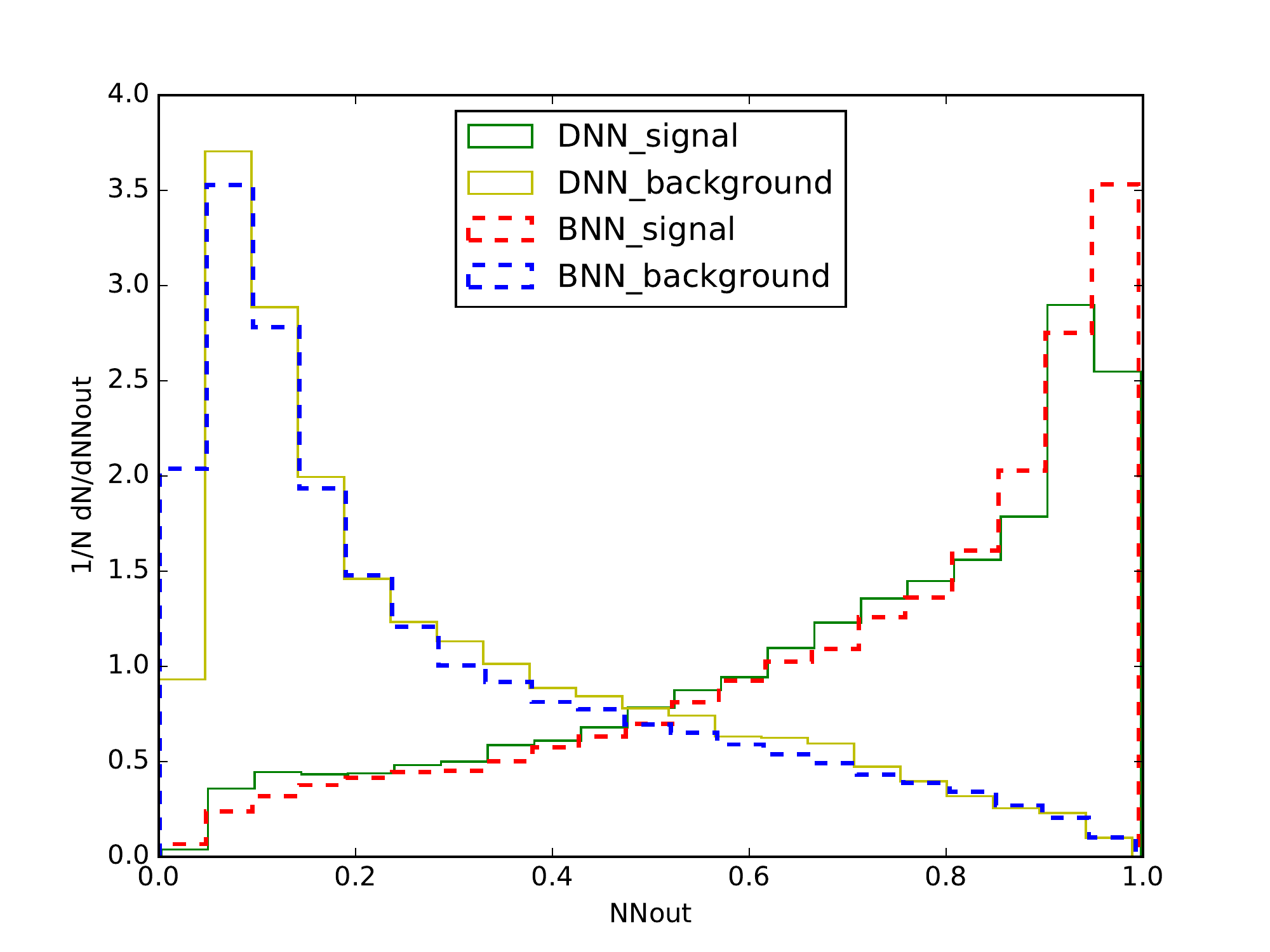}
\end{minipage}
\begin{minipage}[!h!]{.49\linewidth}
  \centering
\includegraphics[width=.95\linewidth]{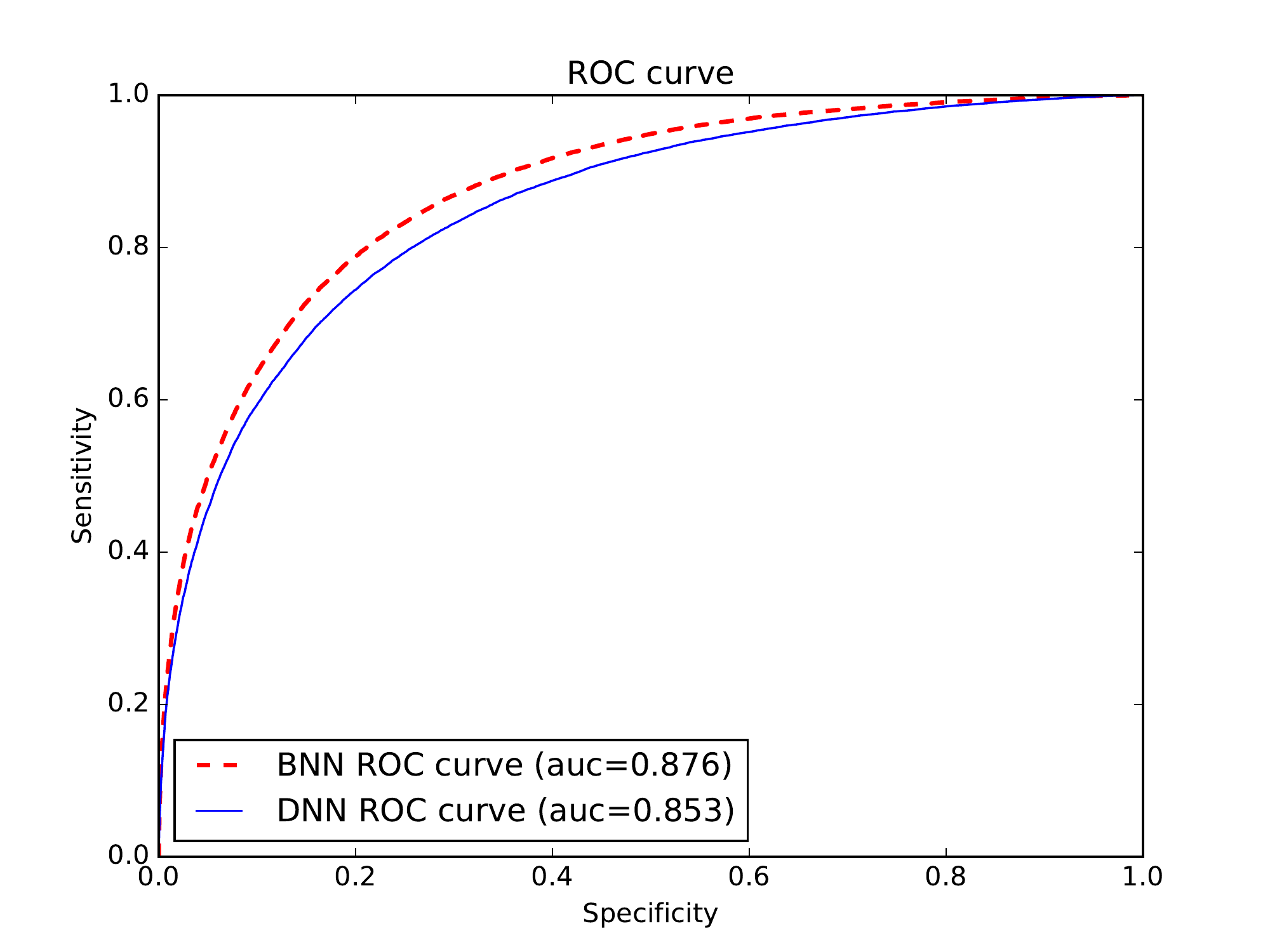}
\end{minipage}
\centering
\caption{\label{low_sp_mand_5layer} The comparison of benchmark BNN with DNN trained on the scalar-products of four-momenta, four-momenta of the final particles and s-channel Mandelstam variables as the set of input variables. DNN has five hidden layers. The left plot demonstrates outputs of DNN and BNN for the signal and background processes. The ROC curves are shown in the right plot.}
\end{figure}

The main properties of the collider hard scattering process, e.g. differential cross sections, are proportional to the squared matrix element of the particular hard process. The concrete form of the matrix elements are different, but in all cases this is a function of scalar products of four-momenta and Mandelstam Lorentz-invariant variables. For example~\cite{Boos:2006xe}, the form of the squared matrix element for the simple s-channel  single-top-quark production process $u\bar{d} \to t\bar{b}$, in terms of scalar products of four-momenta ($p_u,p_b,p_d,p_t$) is the following:
\begin{eqnarray}
|M|^2 = V_{tb}^2 V_{ud}^2 (g_W)^4 
\frac{(p_up_b)(p_dp_t)}{(\hat{s}-m_W^2)^2+\Gamma_W^2 m_W^2} ,
\end{eqnarray}
where $m_W$ and $\Gamma_W$ are the mass and width of W boson, $p$ are the four-momenta of the initial and final particles, $\hat{s}=(p_d+p_t)^2$, and $V_{tb}, V_{ud}, g_W$ are the constants. The matrix element can be rewritten in terms of Mandelstam variables using $(p_up_b) = -\hat{t}/2$ , $(p_dp_t) = (M_t^2 - \hat{t})/2$,
$(p_dp_b) = -\hat{u}/2$ , $(p_up_t) = (M_t^2 - \hat{u})/2$ where $M_t$ is the top quark mass:
\begin{eqnarray}
\label{formula_3}
|M|^2 = V_{tb}^2 V_{ud}^2 (g_W)^4 
\frac{\hat{t} (\hat{t}-M_t^2)}{(\hat{s}-m_W^2)^2+\Gamma_W^2 m_W^2} .
\end{eqnarray}
From the textbooks one knows that for the $2\to n$ scattering processes there are $3n-4$ independent components and minimal set of observables can construct the complete basis. Based on the form of the matrix-element above we would suggest that the correct approach is to take scalar products of four-momenta as the input vector for the DNN analysis. The comparison of such approach with benchmark BNN is shown in Fig.~\ref{sp-3layers}. 
The Fig.~\ref{sp-3layers} demonstrates significantly worse efficiency of DNN trained with scalar products of four-momenta in comparison with benchmark BNN.
The reason is simple in this case, matrix element depends on scalar products of four-momenta  not only of the final particles, but also the four-momenta of the initial particles, which we can not reconstruct for the hadron colliders. 
One can try to compensate the absence of the necessary information with additional observables.  
We add s-channel Mandelstam variables to the set of scalar-products. The results are shown in Fig.~\ref{sp_mand_5layer} the efficiency is better, but still far from the benchmark BNN. In the Fig.~\ref{low_sp_5layer} we compare benchmark BNN with DNN trained on the scalar-products of four-momenta and four-momenta of the final particles as the set of input variables. The efficiency is much better in the last case. For the cross-check, we add s-channel Mandelstam variables to the last set of input variables. In Fig.~\ref{low_sp_mand_5layer} one can see the comparison of benchmark BNN with DNN trained on the scalar-products of four-momenta, four-momenta of the final particles and s-channel Mandelstam variables. The results is practically the same as without s-channel Mandelstam variables.

\begin{figure}[!h!]
\begin{minipage}[!h!]{.49\linewidth}
  \centering
\includegraphics[width=.95\linewidth]{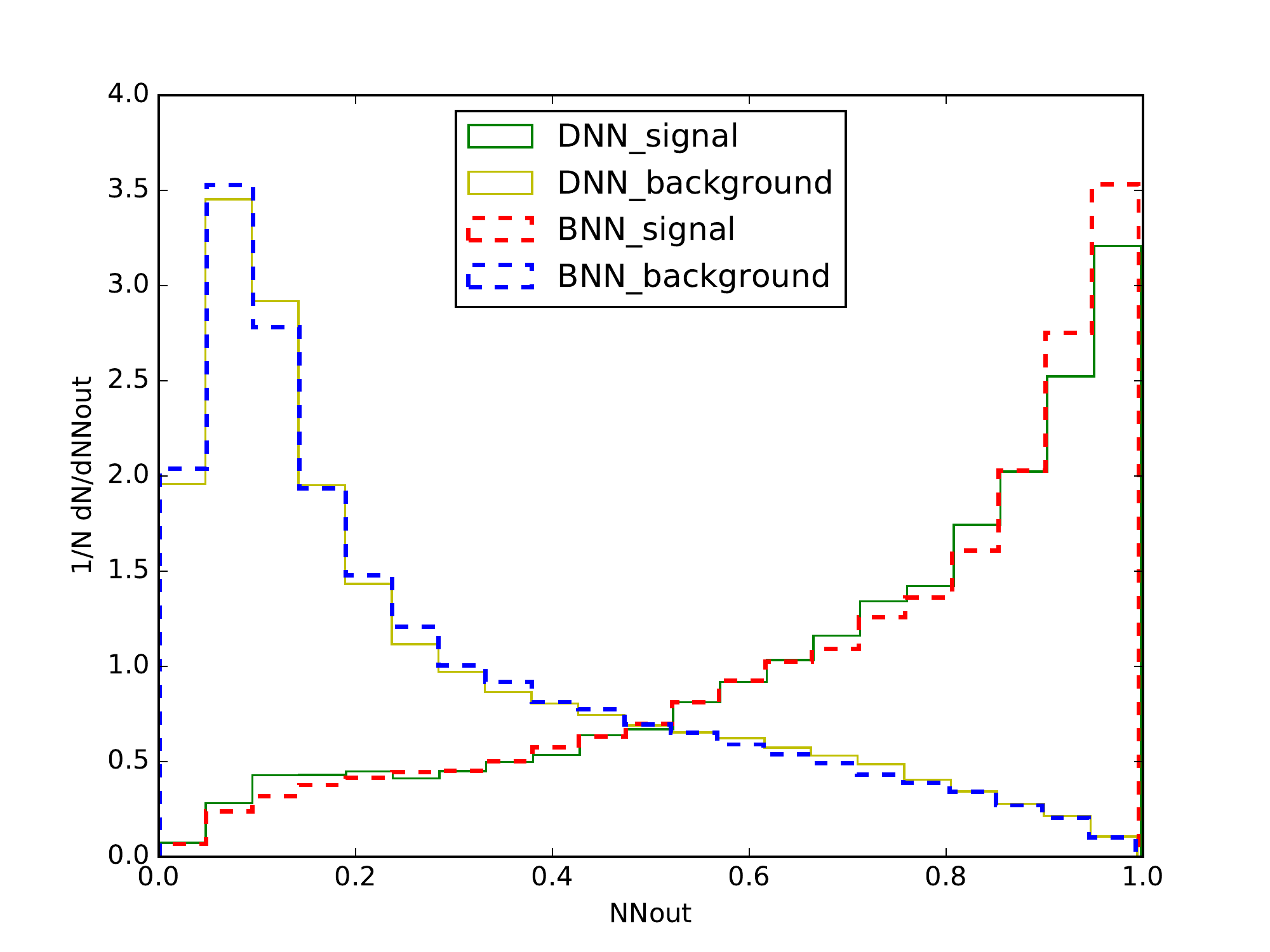}
\end{minipage}
\begin{minipage}[!h!]{.49\linewidth}
  \centering
\includegraphics[width=.95\linewidth]{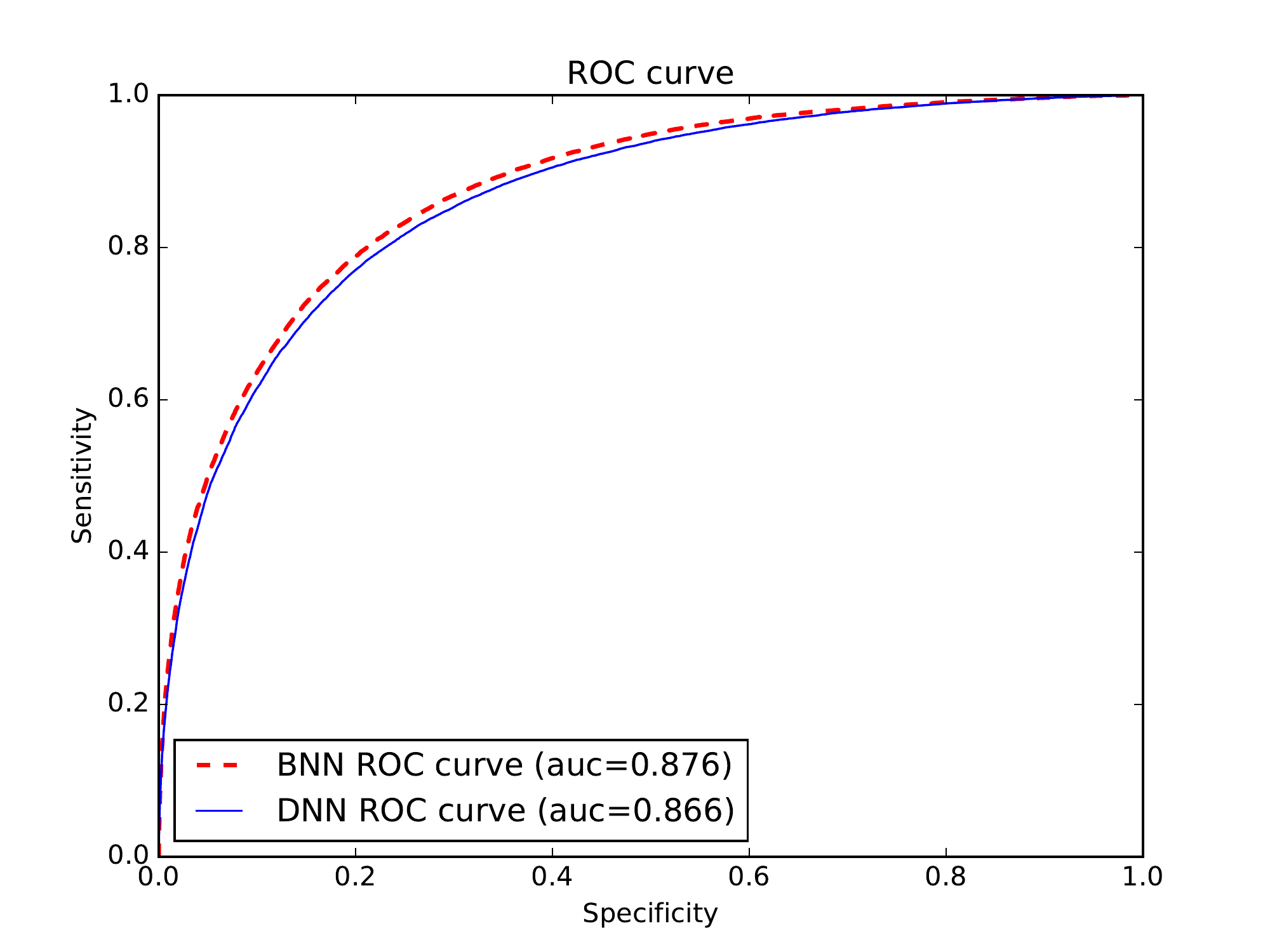}
\end{minipage}
\centering
\caption{\label{pt_sp_mand_3layer} The comparison of benchmark BNN with DNN trained on the scalar-products of four-momenta, transverse momenta of the final particles and Mandelstam variables as the set of input variables. DNN has three hidden layers. The left plot demonstrates outputs of DNN and BNN for the signal and background processes. The ROC curves are shown in the right plot.}
\end{figure}

\begin{figure}[!h!]
\begin{minipage}[!h!]{.49\linewidth}
  \centering
\includegraphics[width=.95\linewidth]{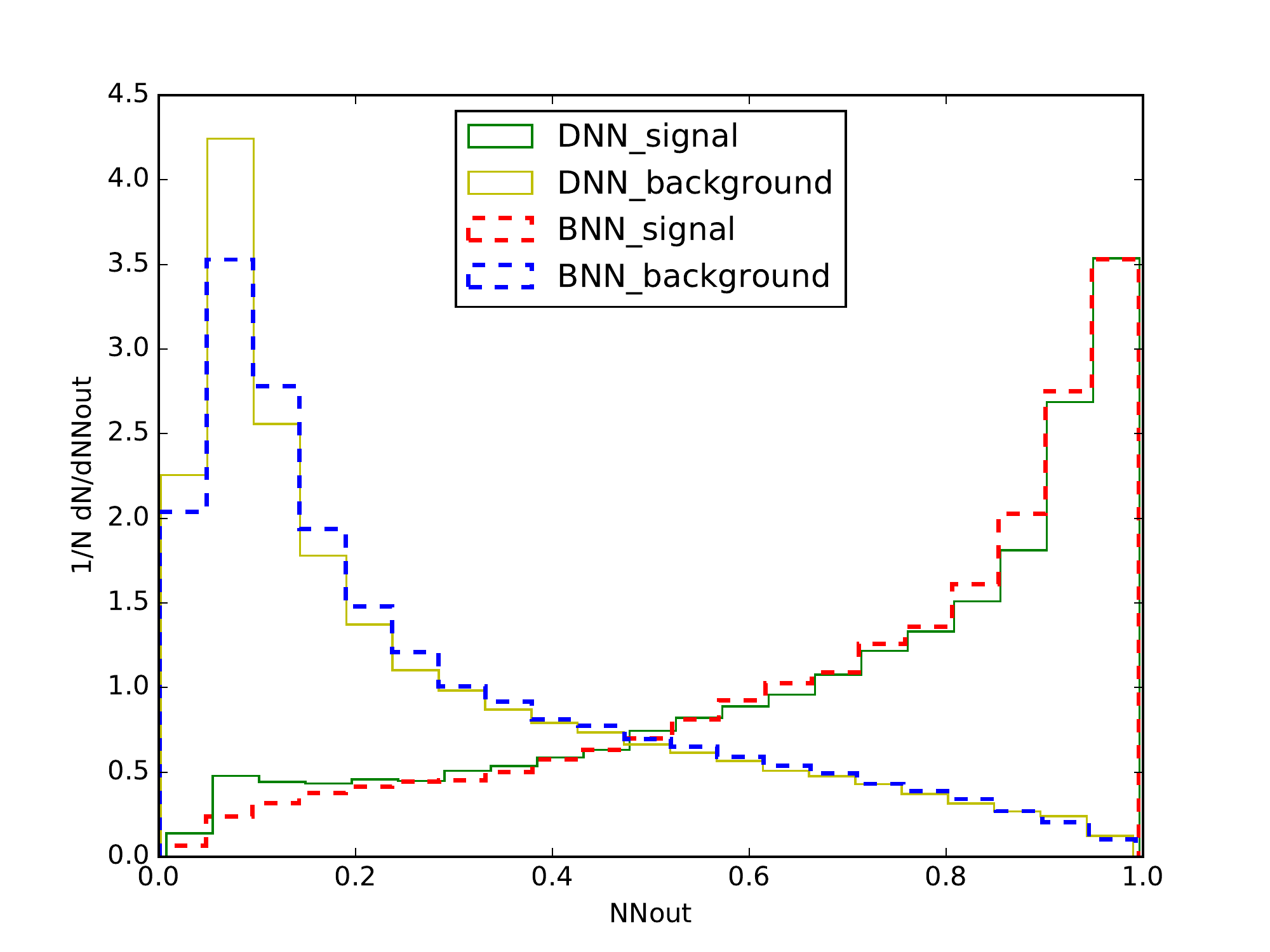}
\end{minipage}
\begin{minipage}[!h!]{.49\linewidth}
  \centering
\includegraphics[width=.95\linewidth]{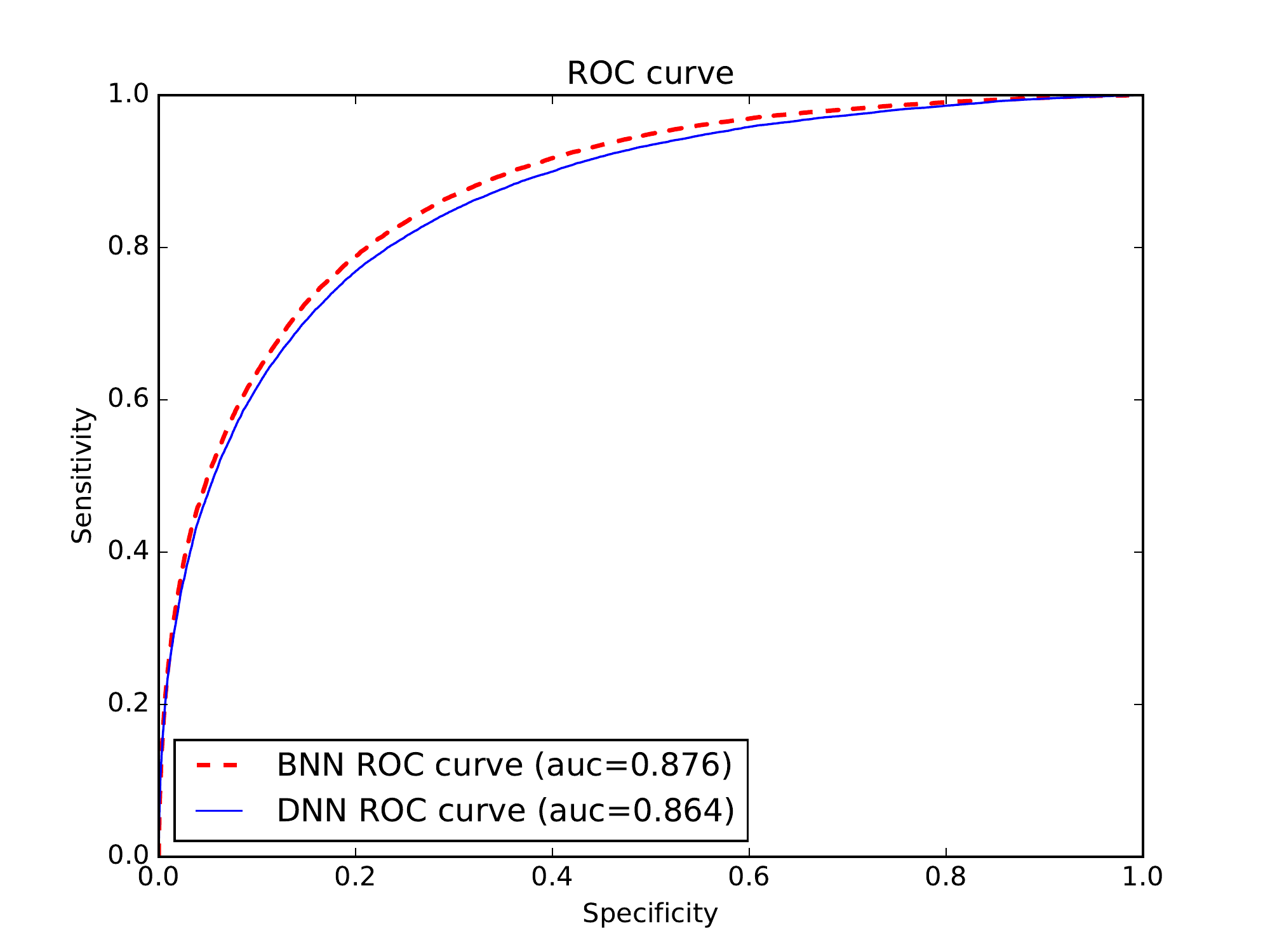}
\end{minipage}
\centering
\caption{\label{pt_low_sp_mand_4layer} The comparison of benchmark BNN with DNN trained on the scalar-products of four-momenta, four-momenta and transverse momenta of the final particles and Mandelstam variables as the set of input variables. DNN has five four layers. The left plot demonstrates outputs of DNN and BNN for the signal and background processes. The ROC curves are shown in the right plot.}
\end{figure}

From the Fig.~\ref{low_sp_5layer} we see that four-momenta of the final particles add significant information and improve efficiency, therefore we need to clarify the reason. 
For the mass-less particles the $t=(p_{\rm final}-p_{\rm initial})^2$ Mandelstam variables can be rewritten in the following form~\cite{Boos:2008sdz}: $\hat{t}_{i,f} = - \sqrt{\hat{s}} e^{Y} p_T^f e^{-|y_f|}$, where $\hat{s}$ is invariant mass of the final particles,  $Y = \frac{1}{2}ln(\frac{p+p_z}{p-p_z})$ is pseudorapidity of the center mass of the system, $P_T^f$ and $y_f$ are transverse momenta and pseudorapidity of the final particle. Therefore, the transverse momenta and pseudorapidity of the final particles can approximate t-channel Mandelstam variables and provide some information on the input particles. On the next step instead of the four-momenta of the final particles we add transverse momenta of the final particles. The result of the comparison is shown in Fig.~\ref{pt_sp_mand_3layer} for the DNN trained on the scalar-products and transverse momenta of the final particles, and s-channel Mandelstam variables. The efficiency is very close to the optimal. For the check we add again four-momenta of the final particles to the input vector. The result is shown in Fig.~\ref{pt_low_sp_mand_4layer} for the DNN trained on the scalar-products, four-momenta and transverse momenta of the final particles, and s-channel Mandelstam variables. We see the absence of additional useful information from the four-momenta of the final particles in comparison with the set of scalar-products and transverse momenta of the final particles, and s-channel Mandelstam variables. 

\begin{figure}[!h!]
\begin{minipage}[!h!]{.49\linewidth}
  \centering
\includegraphics[width=.95\linewidth]{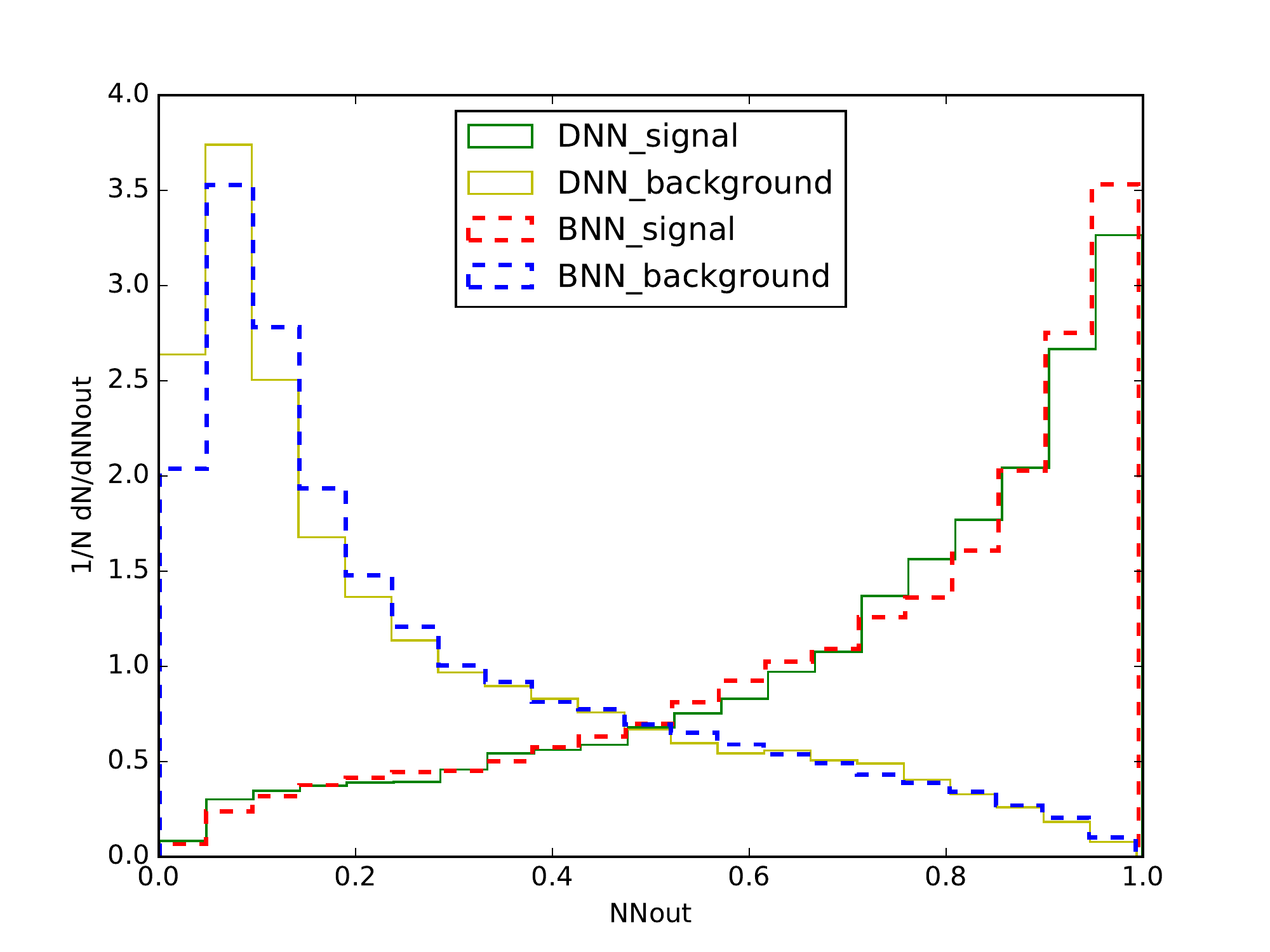}
\end{minipage}
\begin{minipage}[!h!]{.49\linewidth}
  \centering
\includegraphics[width=.95\linewidth]{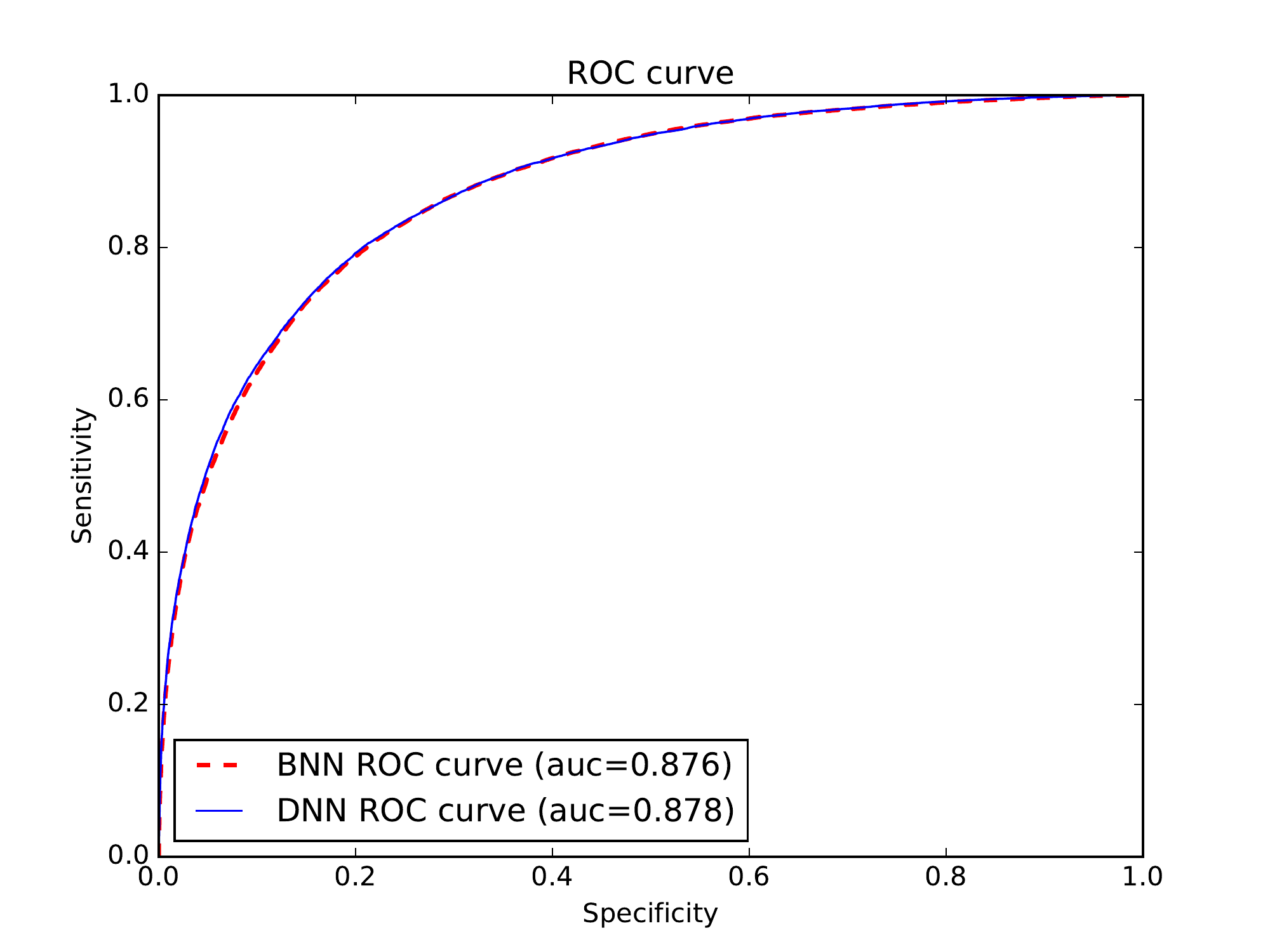}
\end{minipage}
\centering
\caption{\label{high_sp_mand_3layer} The comparison of the benchmark BNN and DNN trained on the proposed general set of low-level variables inspired by the form of  squared  matrix-element. DNN has three hidden layers. The left plot demonstrates outputs of DNN and BNN for the signal and background processes. The ROC curves are shown in the right plot.}
\end{figure}

At the last step of our comparison we add pseudorapidity of the final particles and specific angular variable which is well known for this particular process~\cite{Mahlon:1999gz}. The result is shown in Fig.~\ref{high_sp_mand_3layer}, one can see the comparison of benchmark BNN trained with highly optimized set of high-level variables and DNN trained with the rather general set of low-level variables inspired by the form of  squared  matrix-element, namely the scalar-products of four-momenta of the final particles, transverse momenta and pseudorapidity of the final particles, s-channel Mandelstam variables and specific angular variable. The last angular variable slightly improves the efficiency and is an example of some additional non general information which is specific for the particular task.

\section*{Conclusion}
\label{conclusion}
One of the main advantage of deep learning neural networks is the ability to distinguish sensitive features from raw low-level information.
From the other side, to achieve the optimal performance of DNN one needs to decrease the possible order of nonlinearity in the functional dependence of the input space and desired DNN output. Therefore, one can formulate the general and efficient set of low-level observables 
for the DNN, based on the analysis of the type of functions which represent the necessary properties. 

This paper formulates the general recipe to construct the set of low-level observables for DNN analysis of the hard scattering processes at modern colliders. The recommendation is inspired by the general form of matrix-element square of the hard processes. Several comparisons at different steps are discussed in the paper, demonstrate that the optimal performance can be reached with the following set of low-level observables. First part is the set of scalar-products of four-momenta of the final particles. The second part is the set of transverse momenta and pseudorapidity of the final particles. The third part is the set of s-channel Mandelstam variables. It is also possible to add some specific information which is known for the particular task. It is shown in the paper, one can take such a set of observables for the hard scattering processes at modern colliders and gain close to optimal performance without any sophisticated analysis of the kinematic properties. 

\section*{Acknowledgments}
 The work was supported by grant 16-12-10280 of Russian Science Foundation.

\bibliographystyle{ws-ijmpa}
\bibliography{nnvars}

\begin{thebibliography}{10}
\expandafter\ifx\csname urlstyle\endcsname\relax
  \providecommand{\doi}[1]{doi:\discretionary{}{}{}#1}\else
  \providecommand{\doi}{doi:\discretionary{}{}{}\begingroup
  \urlstyle{rm}\Url}\fi

\bibitem{LevDudko:1999gna}
E.~Boos and L.~Dudko, {\em D0 Collaboration Note D0-3612}   (1999).

\bibitem{Boos:2003gv}
E.~Boos and L.~Dudko, {\em Nucl. Instrum. Meth.} {\bf A502}, 486  (2003),
  \href{http://arxiv.org/abs/hep-ph/0302088}{{\ttfamily arXiv:hep-ph/0302088
  [hep-ph]}}, \doi{10.1016/S0168-9002(03)00477-7}.

\bibitem{Boos:2008sdz}
E.~E. Boos, V.~E. Bunichev, L.~V. Dudko and A.~A. Markina, {\em Phys. Atom.
  Nucl.} {\bf 71}, 388  (2008), \doi{10.1134/s1063778808020191}.

\bibitem{Baldi:2014kfa}
P.~Baldi, P.~Sadowski and D.~Whiteson, {\em Nature Commun.} {\bf 5},   4308
  (2014), \href{http://arxiv.org/abs/1402.4735}{{\ttfamily arXiv:1402.4735
  [hep-ph]}}, \doi{10.1038/ncomms5308}.

\bibitem{Hilbert1}
D.~Hilbert, {\em Bulletin of the American Mathematical Society} {\bf 8}, 461
  (1902).

\bibitem{Hilbert2}
D.~Hilbert, {\em Math. Ann.} {\bf 97}, 243  (1927).

\bibitem{Kolmogorov}
A.~Kolmogorov, {\em Proceedings of the USSR Academy of Sciences} {\bf 108}, 179
   (1956), \doi{10.1090/trans2/017}, [English translation: Amer. Math. Soc.
  Transl., 17 (1961), pp. 369–373.;].

\bibitem{Arnold}
V.~Arnold, {\em Proceedings of the USSR Academy of Sciences} {\bf 114}, 679
  (1957), [English translation: Amer. Math. Soc. Transl., 28 (1963), pp.
  51–54.].

\bibitem{Abadi:2016kic}
M.~Abadi, A.~Agarwal, P.~Barham, E.~Brevdo, Z.~Chen, C.~Citro, G.~S. Corrado,
  A.~Davis, J.~Dean, M.~Devin, S.~Ghemawat, I.~J. Goodfellow, A.~Harp,
  G.~Irving, M.~Isard, Y.~Jia, R.~J{\'{o}}zefowicz, L.~Kaiser, M.~Kudlur,
  J.~Levenberg, D.~Man{\'{e}}, R.~Monga, S.~Moore, D.~G. Murray, C.~Olah,
  M.~Schuster, J.~Shlens, B.~Steiner, I.~Sutskever, K.~Talwar, P.~A. Tucker,
  V.~Vanhoucke, V.~Vasudevan, F.~B. Vi{\'{e}}gas, O.~Vinyals, P.~Warden,
  M.~Wattenberg, M.~Wicke, Y.~Yu and X.~Zheng  (2016),
  \href{http://arxiv.org/abs/1603.04467}{{\ttfamily arXiv:1603.04467}}.

\bibitem{chollet2015keras}
F.~Chollet {\em et~al.}, { Keras} \url{https://github.com/fchollet/keras},
  (2015).

\bibitem{FBMBook}
M.~N. Radford, {\em Bayesian learning for neural networks}, no.~ISBN
  0-387-94724-8 (Dept. of Statistics and Dept. of Computer Science, University
  of Toronto, 1994).

\bibitem{FBMPackage}
M.~N. Radford, {\em {Software for flexible Bayesian modeling and Markov chain
  sampling}}, no.~2004-11-10 (Dept. of Statistics and Dept. of Computer
  Science, University of Toronto, 2004).

\bibitem{Abbott:1999te}
D0 Collaboration, B.~Abbott {\em et~al.}, { {Neural networks for analysis of
  top quark production}}, in {\em {Lepton and photon interactions at high
  energies. Proceedings, 19th International Symposium, LP'99, Stanford, USA,
  August 9-14, 1999}\/},  (1999).
\newblock \href{http://arxiv.org/abs/hep-ex/9907041}{{\ttfamily
  arXiv:hep-ex/9907041 [hep-ex]}}.

\bibitem{Khachatryan:2016sib}
 CMS Collaboration (V.~Khachatryan {\em et~al.}), {\em JHEP} {\bf 02},   028
  (2017), \href{http://arxiv.org/abs/1610.03545}{{\ttfamily arXiv:1610.03545
  [hep-ex]}}, \doi{10.1007/JHEP02(2017)028}.

\bibitem{Boos:2004kh}
 CompHEP Collaboration (E.~Boos, V.~Bunichev, M.~Dubinin, L.~Dudko, V.~Ilyin,
  A.~Kryukov, V.~Edneral, V.~Savrin, A.~Semenov and A.~Sherstnev), {\em Nucl.
  Instrum. Meth.} {\bf A534}, 250  (2004),
  \href{http://arxiv.org/abs/hep-ph/0403113}{{\ttfamily arXiv:hep-ph/0403113
  [hep-ph]}}, \doi{10.1016/j.nima.2004.07.096}.

\bibitem{Sjostrand:2014zea}
T.~Sj{\"o}strand, S.~Ask, J.~R. Christiansen, R.~Corke, N.~Desai, P.~Ilten,
  S.~Mrenna, S.~Prestel, C.~O. Rasmussen and P.~Z. Skands, {\em Comput. Phys.
  Commun.} {\bf 191},   159  (2015),
  \href{http://arxiv.org/abs/1410.3012}{{\ttfamily arXiv:1410.3012 [hep-ph]}},
  \doi{10.1016/j.cpc.2015.01.024}.

\bibitem{deFavereau:2013fsa}
 DELPHES 3 Collaboration (J.~de~Favereau, C.~Delaere, P.~Demin, A.~Giammanco,
  V.~Lemaître, A.~Mertens and M.~Selvaggi), {\em JHEP} {\bf 02},   057
  (2014), \href{http://arxiv.org/abs/1307.6346}{{\ttfamily arXiv:1307.6346
  [hep-ex]}}, \doi{10.1007/JHEP02(2014)057}.

\bibitem{Boos:2006xe}
E.~Boos, V.~Bunichev, L.~Dudko and M.~Perfilov, {\em Phys. Lett.} {\bf B655},
  245  (2007), \href{http://arxiv.org/abs/hep-ph/0610080}{{\ttfamily
  arXiv:hep-ph/0610080 [hep-ph]}}, \doi{10.1016/j.physletb.2007.03.064}.

\bibitem{Mahlon:1999gz}
G.~Mahlon and S.~J. Parke, {\em Phys. Lett.} {\bf B476}, 323  (2000),
  \href{http://arxiv.org/abs/hep-ph/9912458}{{\ttfamily hep-ph/9912458}}.

\end{thebibliography}

\end{document}